\documentclass[a4paper, 12pt]{article}

\pdfoutput = 1

\usepackage{styleBuding}
\usepackage{bm}
\usepackage{color,listings}
\usepackage[usenames,dvipsnames,svgnames,table]{xcolor}
\usepackage{amsthm, amsmath}
\usepackage{amssymb}
\usepackage{mathrsfs}
\usepackage{graphicx}
\usepackage{slashed}
\usepackage{soul}
\usepackage{multirow}
\usepackage{subfigure}
\usepackage{diagbox}
\usepackage{siunitx} 
\usepackage{url} 
\usepackage[utf8]{inputenc}
\usepackage[figuresright]{rotating}
\definecolor{back}{HTML}{F8F8F8}
\usepackage{epstopdf}
\usepackage{lipsum}

\usepackage{booktabs}  
\usepackage{array}  
\usepackage{fancyhdr} 

\preprint{$\begin{gathered}\includegraphics[width=0.05\textwidth]{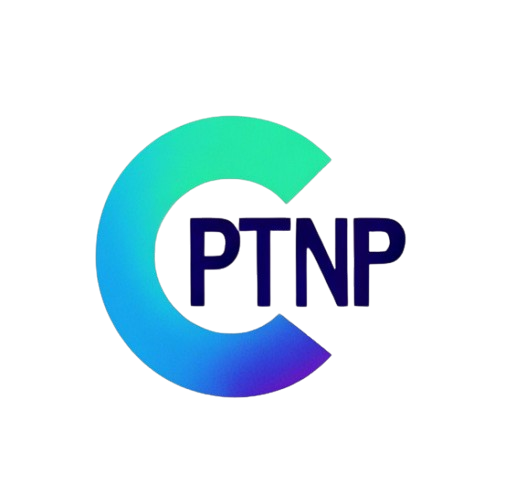}\end{gathered}$\, CPTNP-2025-031}

\newcommand\blfootnote[1]{%
	\begingroup
	\renewcommand\thefootnote{}\footnote{#1}%
	\addtocounter{footnote}{-1}%
	\endgroup
}

\newcommand{\rom}[1]{\uppercase\expandafter{\romannumeral #1\relax}}

\allowdisplaybreaks
\DeclareUnicodeCharacter{2212}{-}

\title{Attractive features of Higgsino Dark Matter in the Next-to-Minimal Supersymmetric Standard Model}
\author{Yuanfang Yue$^a$, Junjie Cao$^{a,b,*}$\blfootnote{*Corresponding author.}, Fei Li$^a$, and Zehan Li$^a$}
\affiliation{ $^a$ School of Physics, Henan Normal University, Xinxiang 453007, China}
\affiliation{ $^b$ School of Physics, Zhengzhou University, Zhengzhou 450000, China}
\emailAdd{yueyuanfang@htu.edu.cn}
\emailAdd{junjiec@alumni.itp.ac.cn}
\emailAdd{hnufeili@163.com}
\emailAdd{hnuzehanli@163.com}
\abstract{In the Higgsino dark matter (DM) scenario of the Minimal Supersymmetric Model (MSSM), the mixing of Gaugino and Higgsino influences the mass splitting between neutralinos predominantly composed of Higgsino and introduces coupling between the DM and Higgs bosons. These effects modify the DM-nucleon scattering cross-section, causing conflicts with the latest direct detection results from LZ experiments for both substantial and minute mixings. Consequently, the experimental measurement of DM relic density necessitates the Higgsino DM mass to be approximately 1.1 TeV. We discovered that in the Higgsino DM scenario of the Next-to-Minimal Supersymmetric Model (NMSSM), the mixing of Higgsino and Singlino introduces analogous effects, with a crucial distinction being that the current LZ experiment permits significant mixing between Singlino and Higgsino. This pronounced mixing effect effectively attenuates the interactions between Higgsino-dominated neutralinos and standard model particles, enabling DM masses exceeding roughly 670 GeV to achieve the correct relic abundance. Through analytical formulas and numerical results, we elucidated these characteristics which were not observed before. Our research reveals that in the NMSSM, when comprehensively examining the mixing effects of Higgsino, Gaugino, and Singlino, the properties of Higgsino DM become markedly more intricate compared to the MSSM predictions.}

\begin{document}
    \maketitle
    \flushbottom

\section{Introduction}

The existence of dark matter (DM) has been firmly established by various astronomical and cosmological observations~\cite{Planck:2018vyg}, yet its particle nature remains one of the most profound mysteries in modern physics. Among various theoretical candidates, the Higgsino, which naturally emerges from supersymmetric theories, stands out as a particularly compelling DM candidate~\cite{1995Supersymmetric,Kane:1996dd,dreesLightHiggsinoDark1997,Masip:2005fv,Wang:2005kf,Hall:2009aj,Sinha:2012kq,Mayes:2014wda,Kowalska:2018toh,Delgado:2020url,Wang:2024ozr,Evans:2022gom}. In the Minimal Supersymmetric Standard Model (MSSM)~\cite{Nilles:1983ge,Haber:1984rc,Gunion:1984yn,Martin:1997ns}, the Higgsino has 
the following distinctive features that make it an attractive solution to the DM puzzle, drawing on both theoretical predictions and experimental constraints:

\begin{itemize}
\item From a theoretical perspective, the Higgsino possesses remarkable simplicity and naturalness~\cite{Ellwanger:2018zxt,Baer:2018rhs,Bae:2017hlp,baerRadiativeNaturalSupersymmetry2013a,Han:2019vxi}. As an SU(2) doublet, it represents one of the most minimal extensions to the Standard Model (SM) that can accommodate viable DM candidates~\cite{Aoki:2011he,Bharucha:2017ltz}. This theoretical elegance is further enhanced by its crucial role in gauge coupling unification, where the Higgsino provides necessary contributions to achieve precise unification at high energies~\cite{Kowalska:2014hza,Aparicio:2016qqb,SuryanarayanaMummidi:2020ydm} and is readily embedded into theories like split SUSY~\cite{Fox:2014moa,Arkani-Hamed:2004ymt,Giudice:2004tc,Arkani-Hamed:2004zhs}. Additionally, in the MSSM, Higgsino DM depends only on a limited number of parameters, minimizing theoretical ambiguities and providing a clean framework for both theoretical study and experimental interpretation~\cite{Kowalska:2018toh,Martin:2024pxx}.
    
\item A particularly intriguing aspect of Higgsino DM is its mass scale. To produce the observed DM relic density, the Higgsino mass needs to be around 1.1 TeV~\cite{Chattopadhyay:2005mv,Chakraborti:2014fha,Fox:2014moa,Delgado:2020url,Shafi:2023sqa,Mummidi:2018myd}. This mass scale, determined by the interplay between electroweak interactions and freeze-out dynamics, carries profound physical significance. Specifically, it is not only preferred by natural electroweak symmetry breaking, as indicated by the expression of the Z boson mass~\cite{Baer:2012uy}, but also explains the current null results from the Large Hadron Collider (LHC) searches for supersymmetry~\cite{ATLAS:2024exu}. The latter aspect is characterized by the fact that the LHC's sensitivity to both electroweakly and strongly produced supersymmetric particles decreases monotonously as the DM becomes heavy, making it challenging to detect their signals when the DM mass is larger than approximately $1~{\rm TeV}$~\cite{ATLAS:2024exu}. This alignment between theoretical prediction and experimental observation may be viewed as a compelling hint for the Higgsino DM scenario.
    
\item The phenomenology of Higgsino DM exhibits rich features in terms of experimental detection~\cite{Chen:2019gtm,Hicyilmaz:2020bph,Wang:2020xta,Chun:2016cnm}. In direct detection experiments, the interaction of Higgsino DM with nuclei depends on its mixing with heavier Gaugino states (e.g., Wino or Bino). When Gaugino component mixing increases, the scattering cross-section becomes large~\cite{Yanagida:2013uka,Co:2021ion,Rinchiuso:2020skh,Bisal:2023fgb}, rendering such models stringently constrained by leading DM direct detection experiments such as PandaX-4T~\cite{PandaX-4T:2021bab}, XENONnT~\cite{XENON:2023cxc}, and LUX-ZEPLIN (LZ)~\cite{LZ:2022lsv,LZ:2024zvo}. Current observations suggest that Gaugino masses must be larger than around 2 TeV to avoid over-enhancing the Higgsino's spin-independent (SI) scattering cross-section~\cite{Co:2021ion,Nagata:2014wma,Martin:2024pxx,Martin:2024ytt}.

Furthermore, when the Gauginos are heavier than $10^8 ~{\rm GeV}$, the mass difference between the lightest neutral Higgsino components is smaller than 100 KeV~\cite{Chun:2016cnm}. In this case,  inelastic DM-nucleon scattering processes in direct detection are kinematically allowed with an unsuppressed spin-dependent (SD) cross-section, which has been very tightly constrained by current detection experiments. These features yield a unique phenomenological window for the Higgsino, where the Gauginos with masses between 2 TeV and $10^8 \text{GeV}$ naturally evade current experimental bounds. 

In addition, unlike Wino DM which acts as another popular candidate, Higgsino DM can avoid indirect detection constraints by lacking significant Sommerfeld enhancement in its annihilation cross-section due to the larger mass splitting between the charged and neutral components of the Higgsino field~\cite{dessert2023higgsino,Rodd:2024qsi,Beneke:2016jpw}. This makes it challenging for current gamma-ray~\cite{Chen:2010yi,Beneke:2019gtg} or neutrino telescopes~\cite{Enberg:2015qwa} to detect Higgsino annihilation signals, adding to the difficulty in probing this candidate with indirect detection experiments~\cite{dessert2023higgsino}.

\item Higgsino DM exhibits unique collider signatures due to the small mass splittings among its charged and neutral components. The charged Higgsino (chargino) can manifest as either missing transverse energy (MET) or disappearing tracks in high-energy collisions~\cite{Bobrovskyi:2012dc}. These signals are a robust prediction of electroweak-scale Higgsino DM and present both challenges and opportunities for future experiments~\cite{Chakraborti:2024pdn,Chakraborti:2017dpu,Saito:2019rtg,Capdevilla:2021fmj,Delgado:2020url}. Advanced detection techniques, such as optimized triggers for soft visible decay products and improved track reconstruction algorithms, are currently being developed, making this an active area of theoretical and experimental research~\cite{Fukuda:2017jmk,Martin:2024pxx}
\end{itemize}

In conclusion, Higgsino DM is regarded as one of the most theoretically elegant and experimentally compelling candidates for addressing the DM enigma. Its profound connection to supersymmetry, compatibility with current experimental constraints, and distinctive phenomenological characteristics have established it as a central focus of ongoing research.

While the MSSM offers a robust theoretical framework for studying Higgsino DM, it faces the challenging $\mu$-parameter problem. Although this issue can be addressed through the well-known Giudice-Masiero mechanism~\cite{Giudice:1988yz}, implementing this solution introduces significant fine-tuning in electroweak symmetry breaking when considering the LHC's supersymmetry search results and the DM experimental detection results~\cite{Arvanitaki:2013yja,Evans:2013jna,Baer:2014ica,Cao:2019qng}.

To address these challenges, we extend our investigation of Higgsino DM to the Next-to-Minimal Supersymmetric Standard Model (NMSSM), which serves as a natural and minimal extension of the MSSM~\cite{Miller:2003ay,Ellwanger:2009dp,Maniatis:2009re}.   This model incorporates an additional gauge singlet Higgs superfield, $\hat{S}$, that can dynamically generate an effective $\mu$ parameter when its scalar component develops a vacuum expectation value (vev).  
This mechanism substantially improves the stability of the electroweak vacuum, making the theoretical framework more appealing~\cite{Mummidi:2018myd,Hollik:2018yek,Hollik:2018wrr}.

In the NMSSM, the fermionic component of $\hat{S}$, known as the Singlino, can substantially mix with the Higgsino field, thereby altering key properties of Higgsino DM, such as its mass splitting from other Higgsino-like particles and its interactions with nuclei~\cite{Cao:2016nix,Mummidi:2018myd}.  Consequently, Higgsino DM exhibits  novel features in DM annihilation and DM-nucleon scattering processes. 

Given that these characteristics have been scarcely discussed in previous literatures, this study aims to comprehensively examine them. Notably, in the $Z_3$-invariant NMSSM, achieving substantial Higgsino-Singlino mixing necessitates near-degeneracy of the fields' masses, constraining the Yukawa couplings $\lambda$ and $\kappa$ to satisfy $\lambda \simeq 2 \kappa$~\cite{Ellwanger:2009dp}. As will be illustrated below, this constraint limits the properties of Higgsino DM. Therefore, we conduct our investigation within the framework of the General NMSSM (GNMSSM)~\cite{Cao:2022ovk,Cao:2024axg,Meng:2024lmi} to provide a more complete characterization of these properties. We will show that the properties of the Higgsino DM in the NMSSM are more flexible than those in the MSSM, while remaining consistent with various experimental constraints.

The remainder of this paper is structured as follows. Section~\ref{theory-section} delineates the distinctive characteristics of the GNMSSM, with a particular focus on the analytical examination of the Higgsino DM properties through expressions of various observables. Section~\ref{numerical study} elucidates the research methodology and numerical outcomes that highlight the novel aspects of the Higgsino DM. Section~\ref{Other_Issues} discusses the context of our study in relation to previous work and its prospects for future detection and other related issues. Finally, Section~\ref{conclusion} offers a comprehensive summary of the key research findings.

\section{\label{theory-section}Theoretical preliminaries} 
\subsection{\label{Section-Model}Basics of the GNMSSM }

The GNMSSM extends the MSSM by introducing a gauge singlet superfield $\hat{S}$, which does not carry any baryonic or leptonic number. Consequently, its Higgs sector comprises $\hat{S}$ alongside two $SU(2)_L$ doublet superfields,  $\hat{H}_u=(\hat{H}_u^+,\hat{H}_u^0)$ and $\hat{H}_d=(\hat{H}_d^0,\hat{H}_d^-)$. The general form of the GNMSSM superpotential is given by~\cite{Ellwanger:2009dp}:
\begin{eqnarray}
W_{\rm GNMSSM} = W_{\text{Yukawa} }+ \lambda \hat{S}\hat{H_u} \cdot \hat{H_d} + \frac{\kappa}{3}\hat{S}^3 + \mu \hat{H_u} \cdot \hat{H_d}  + \frac{1}{2} \mu^{\prime} \hat{S}^2 + \xi\hat{S}, \label{Superpotential}
\end{eqnarray}
where $W_{\rm Yukawa}$ contains the quark and lepton Yukawa interactions from the MSSM superpotential but excludes the $\mu$-term, and the dimensionless couplings $\lambda$ and $\kappa$ parameterize the Higgs-sector interactions, similar to the $Z_3$-NMSSM. The parameters $\mu$, $\mu^\prime$, and $\xi$ describe $Z_3$-symmetry-violating effects, which help address the tadpole problem~\cite{Ellwanger:1983mg, Ellwanger:2009dp} and the cosmological domain-wall problem in the $Z_3$-NMSSM~\cite{Abel:1996cr, Kolda:1998rm, Panagiotakopoulos:1998yw}. Since one of these parameters can be eliminated by shifting $\hat{S}$ with a constant and redefining the other parameters~\cite{Ross:2011xv}, we set $\xi=0$ without loss of generality.
Previous studies indicate that $\mu$ and $\mu^\prime$ can naturally lie in the range of a few hundred GeV from the breaking of the fundamental discrete $R$-symmetry $\mathbb{Z}^R_4$ or $\mathbb{Z}^R_8$ at high energies~\cite{Abel:1996cr, Lee:2010gv, Lee:2011dya, Ross:2011xv, Ross:2012nr}. These parameters can substantially modify the properties of neutral Higgs bosons and neutralinos, leading to a richer phenomenology than those in the $Z_3$-NMSSM and MSSM, which forms the main focus of this study.

In contrast to the MSSM, the GNMSSM contains both the traditional $\mu$-term $\mu \hat{H}_u \cdot \hat{H}_d$ and a dynamically generated $\mu$-term $\lambda \hat{S} \hat{H}_u \cdot \hat{H}_d$. When the scalar component of $\hat{S}$ develops a vev $\langle S \rangle \equiv v_s/\sqrt{2}$, the effective $\mu$ parameter is given by $\mu_{\rm tot} = \lambda v_s/\sqrt{2} + \mu$. This dynamical generation provides additional flexibility in addressing the naturalness problems associated with the $\mu$-term in the MSSM. In particular, if $\lambda v_s/\sqrt{2} \gg \mu$, $\mu_{\rm tot}$ is predominantly generated dynamically.

\subsection{\label{DMRD}The Higgs Sector}

In the GNMSSM, the soft SUSY-breaking terms in the Higgs sector include trilinear interactions involving the singlet and doublet fields, as well as conventional mass terms. These terms are expressed as:
\begin{eqnarray}
    -\mathcal{L}_{soft} = &\Bigg[\lambda A_{\lambda} S H_u \cdot H_d + \frac{1}{3} \kappa A_{\kappa} S^3+ m_3^2 H_u\cdot H_d + \frac{1}{2} {m_S^{\prime}}^2 S^2 + \xi^\prime S + h.c.\Bigg]  \nonumber \\
   & + m^2_{H_u}|H_u|^2 + m^2_{H_d}|H_d|^2 + m^2_{S}|S|^2 , \label{Soft-terms}
     \end{eqnarray}
After electroweak symmetry breaking, the CP-even and CP-odd Higgs fields mix to form three and two mass eigenstates, respectively. In this context, it is more intuitive to work with physical parameters rather than the original Lagrangian parameters. The key physical parameters are given as the follows~\cite{Meng:2024lmi}:  
\begin{itemize}
\item $m_A$: This parameter represents the mass scale of the heavy MSSM-like CP-odd Higgs boson. It is defined as
\begin{eqnarray}
m_A^2 = \left [ \lambda v_s (\sqrt{2} A_\lambda + \kappa v_s + \sqrt{2} \mu^\prime ) + 2 m_3^2 \right ]/\sin 2 \beta. \label{mA}
\end{eqnarray}
\item $m_B$: This parameter characterizes the mass scale of the CP-even singlet Higgs. It is related to the original parameters by
\begin{eqnarray}
m_B^2 &=& \frac{(A_\lambda + \mu^\prime) \sin 2 \beta}{2 \sqrt{2} v_s} \lambda v^2   + \frac{\kappa v_s}{\sqrt{2}} (A_\kappa +  2 \sqrt{2} \kappa v_s + 3 \mu^\prime ) \nonumber \\
& & - \frac{\mu}{\sqrt{2} v_s} \lambda v^2 - \frac{\sqrt{2}}{v_s} \xi^\prime.  
\end{eqnarray}
\item $m_C$: This parameter describes the mass scale of the CP-odd singlet Higgs, determined by 
\begin{eqnarray}
m_C^2 &=&\frac{(A_\lambda + 2 \sqrt{2} \kappa v_s + \mu^\prime ) \sin 2 \beta }{2 \sqrt{2} v_s} \lambda v^2  - \frac{\kappa v_s}{\sqrt{2}} (3 A_\kappa + \mu^\prime)  \nonumber \\
& & - \frac{\mu}{\sqrt{2} v_s} \lambda v^2 - 2 m_S^{\prime\ 2} - \frac{\sqrt{2}}{v_s} \xi^\prime. \label{mC}
\end{eqnarray}
\item $m_N$ and $\mu_{\rm tot}$: These denote the Singlino and the Higgsino masses, respectively, and are given by
\begin{eqnarray}
m_N = \mu^\prime + \sqrt{2} \kappa v_s, \quad \mu_{\rm tot} = \mu + \lambda v_s/\sqrt{2}. \label{mN} 
\end{eqnarray}
\end{itemize}

In the basis defined by ($H_{\rm NSM} \equiv \cos\beta {\rm Re}(H_u^0) - \sin\beta {\rm Re}(H_d^0)$, $H_{\rm SM} \equiv \sin\beta {\rm Re}(H_u^0) + \cos\beta {\rm Re} (H_d^0)$, ${\rm Re[S]}$), elements of the CP-even mass matrix read as~\cite{Meng:2024lmi}
\begin{eqnarray}
{\cal M}^2_{S, 11}&=& m_A^2 + \frac{1}{2} (2 m_Z^2- \lambda^2v^2)\sin^22\beta, \quad {\cal M}^2_{S, 12}=-\frac{1}{4}(2 m_Z^2-\lambda^2v^2)\sin4\beta, \nonumber \\
  {\cal M}^2_{S, 13} &=& -\frac{\lambda v}{\sqrt{2}} ( A_\lambda + m_N ) \cos 2 \beta, \quad {\cal M}^2_{S, 22} = m_Z^2\cos^22\beta+ \frac{1}{2} \lambda^2v^2\sin^22\beta, \nonumber  \\
  {\cal M}^2_{S, 23}&=& \frac{\lambda v}{\sqrt{2}} \left[ 2 \mu_{tot} - (A_\lambda + m_N ) \sin2\beta \right], \quad {\cal M}^2_{S, 33} = m_B^2.  \label{CP-even-mass-matrix}
\end{eqnarray}
Similarly, in the basis ($A_{\rm NSM} \equiv \cos\beta {\rm Im}(H_u^0) - \sin\beta  {\rm Im}(H_d^0)$, ${\rm Im[S]}$), the CP-odd mass matrix elements are~\cite{Meng:2024lmi}      
\begin{eqnarray}
{\cal M}^2_{P,11} = m_A^2, \quad {\cal M}^2_{P,22} &=& m_C^2, \quad {\cal M}^2_{P,12} = \frac{\lambda v}{\sqrt{2}} ( A_\lambda - m_N ).  \label{CP-odd-mass-matrix}
\end{eqnarray}   
Here, $\tan \beta$ is the ratio of the Higgs doublet vevs, i.e., $\tan \beta \equiv v_u/v_d$, with $v = \sqrt{v_u^2+v_d^2}\simeq 246~\mathrm{GeV}$.

Diagonalizing these matrices yields the mass eigenstates $h_i=\{h, H, h_s\}$ and $a_j=\{A_H, A_s\}$, which are related to the interaction states via 
\begin{eqnarray} \label{Mass-eigenstates}
  h_i & = & V_{h_i}^{\rm NSM} H_{\rm NSM}+V_{h_i}^{\rm SM} H_{\rm SM}+V_{h_i}^{\rm S} Re[S], \\
  a_j & = &  V_{P, a_j}^{\rm NSM} A_{\rm NSM}+ V_{P, a_j}^{\rm S} Im [S]. \label{higgsMix}
\end{eqnarray} 
In this notation,  $h$ corresponds to the scalar discovered at the LHC, $H$ and $A_H$ represent heavy doublet-dominated Higgs bosons and $h_s$ and $A_s$ are singlet-dominated states. For convenience, these states are also labeled as $h_i$ (i=1,2,3) and $A_j$ (j=1,2) in ascending mass orders, i.e., $m_{h_1} < m_{h_2} < m_{h_3}$ and $m_{A_1} < m_{A_2}$ in this study. 
The model also predicts a pair of charged Higgs bosons, $H^\pm = \cos \beta H_u^\pm + \sin \beta H_d^\pm$, with masses given by~\cite{Ellwanger:2009dp}
\begin{eqnarray}
    m^2_{H^{\pm}} &=&  m_A^2 + m^2_W - \frac{1}{2}\lambda^2 v^2. \label{Charged Hisggs Mass}
  \end{eqnarray}

Several important features characterize the GNMSSM Higgs sector:
\begin{itemize}
\item Higgs mass spectrum 

The LHC data reveal that one eigenstate must exhibit SM-like couplings and have a mass of approximately 125 GeV~\cite{ATLAS:2022vkf,CMS:2022dwd}. Additionally, the doublet-like states should be heavier than approximately $1~{\rm TeV}$, while singlet-dominated states are less tightly restricted~\cite{ATLAS:2020zms,ATLAS:2021upq}. 
\item Doublet-Singlet mixing

The mixing of the singlet states with doublet fields is proportional to $\lambda$.  In the decoupling limit $\lambda \to 0$, the singlet fields effectively separate from the Higgs doublets, and the masses $m_B$, $m_C$, and $m_N$ can be treated as physical particle masses with high accuracy.

\item Heavy charged Higgs approximation

When the charged Higgs bosons are heavy, they are approximately degenerate in mass with the CP-even scalar $H$ and the CP-odd scalar $A_H$. Simplified expressions for the singlet-dominated states and their mixing parameters can be used in this limit~\cite{Baum:2017enm}:
\begin{eqnarray}
m_{h_s}^2 & \simeq & m_B^2 - \frac{{\cal M}^4_{S, 13}}{m_A^2 - m_B^2}, \quad m_{A_s}^2 \simeq m_C^2 - \frac{{\cal M}^4_{P, 12}}{m_A^2 - m_C^2}, \quad \frac{V_{P,A_s}^{\rm NSM}}{V_{P,A_s}^{\rm S}} = \frac{{\cal M}^2_{P, 12}}{m_{A_s}^2 - m_A^2} \simeq 0, \nonumber \\
\frac{V_{h}^{\rm S}}{V_h^{\rm SM}} & \simeq &  \frac{{\cal M}^2_{S, 23}}{m_h^2 - m_B^2}, \quad V_{h}^{\rm NSM} \sim 0, \quad V_h^{\rm SM} \simeq \sqrt{1 - \left ( \frac{V_{h}^{\rm S}}{V_h^{\rm SM}} \right )^2}  \sim 1, \nonumber  \\
\frac{V_{h_s}^{\rm SM}}{V_{h_s}^{\rm S}} & \simeq &  \frac{{\cal M}^2_{S, 23}}{m_{h_s}^2 - m_h^2}, \quad V_{h_s}^{\rm NSM} \sim 0, \quad V_{h_s}^{\rm S} \simeq \sqrt{1 - \left ( \frac{V_{h_s}^{\rm SM}}{V_{h_s}^{\rm S}} \right )^2 } \sim 1.   \label{Approximations}
\end{eqnarray}
These formulae indicate that $V_h^S \simeq - V_{h_s}^{\rm SM}$ and they are all proportional to $\lambda$. 
\item Parameter determination

Using the physical masses $m_A$, $m_B$, $m_C$, $m_N$, and $\mu_{\rm tot}$ as inputs, the original Lagrangian parameters can be determined~\cite{Meng:2024lmi}: 
\begin{eqnarray}
\mu &= & \mu_{tot} - \frac{\lambda}{\sqrt{2}} v_s, \quad \mu^\prime = m_N - \sqrt{2} \kappa v_s, \nonumber \\
m^2_3 &= & \frac{m^2_A \sin{2\beta}}{2} - \lambda v_s (\frac{\kappa v_s}{2} + \frac{\mu^\prime}{\sqrt{2}} + \frac{A_\lambda}{\sqrt{2}}), \nonumber \\
\xi^\prime &=& \frac{v_s}{\sqrt{2}} \left [ \frac{(A_\lambda + \mu^\prime) \sin 2 \beta}{2 \sqrt{2} v_s} \lambda v^2   + \frac{\kappa v_s}{\sqrt{2}} (A_\kappa +  2 \sqrt{2} \kappa v_s + 3 \mu^\prime ) - \frac{\mu}{\sqrt{2} v_s} \lambda v^2 - m_B^2 \right ],  \nonumber\\
m_S^{\prime 2} &=& \frac{1}{2} \left[ m_B^2 - m^2_C + \lambda \kappa \sin 2\beta v^2 - 2 \sqrt{2} \kappa v_s ( A_\kappa + \frac{\kappa}{\sqrt{2}} v_s + \mu^\prime) \right].   \label{Simplify-1}
\end{eqnarray}
  
\item Parameter dependence

In the Higgs sector, eight out of eleven parameters, namely $\tan \beta$, $\lambda$, $A_\lambda$, $m_A$, $m_B$, $m_C$, $m_N$, and $\mu_{tot}$, uniquely determine the Higgs mass matrices. The remaining three parameters, $\kappa$, $A_\kappa$, and $v_s$, play a distinct role in governing the triple Higgs coupling strengths~\cite{Meng:2024lmi}. In the vanishing tadpole scenario delineated by $\xi^\prime = 0$, $A_\kappa$ takes the form~\cite{Meng:2024lmi}:
\begin{eqnarray}
\kappa A_\kappa & = & \frac{\sqrt{2} m^2_B}{v_s} + \frac{\lambda \mu v^2}{v_s^2} - \frac{ \lambda (A_\lambda + m_N - \sqrt{2} \kappa v_s) v^2  \sin{2\beta}}{2 v_s^2} \nonumber \\
& & + \sqrt{2} \kappa^2 v_s - 3 \kappa m_N.  
\end{eqnarray}
Moreover, investigations of the light $h_s$ scenario usually employ a parameterization of  ${\cal M}^2_{S, 23}$ as ${\cal M}^2_{S, 23} = \sqrt{2} \lambda \delta v \mu_{\rm tot}$, where $\delta$ is defined by
$\delta \equiv [ 2 \mu_{\rm tot} - (A_\lambda + m_N) \sin 2 \beta]/(2 \mu_{\rm tot})$~\cite{Lian:2024smg}.
This parameter $\delta$ serves as a measure of the cancellation between different terms in ${\cal M}^2_{S, 23}$. A key advantage of this parameterization emerges when $\delta$ takes small values: it naturally accommodates larger values of $\lambda$ while maintaining consistency with the LHC measurements of the $125~{\rm GeV}$ Higgs properties.
\end{itemize}

\subsection{Neutralino Sector}

The mixing between the fermionic partners for the neutral Higgs bosons and the gauginos gives rise to five neutralinos and two charginos, denoted by $\tilde{\chi}_i^0$ ($i=1,\dots,5$) and $\tilde{\chi}_i^\pm$ ($i=1,2$), respectively.
In the gauge eigenstate basis $\psi^0 = \left(-i \tilde{B}, -i\tilde{W}, \tilde{H}_d^0, \tilde{H}_u^0, \tilde{S}\right)$, we can express the symmetric neutralino mass matrix as~\cite{Ellwanger:2009dp}
\begin{equation}
    M_{\tilde{\chi}^0} = \left(
    \begin{array}{ccccc}
    M_1 & 0 & -m_Z \sin \theta_W \cos \beta & m_Z \sin \theta_W \sin \beta & 0 \\
      & M_2 & m_Z \cos \theta_W \cos \beta & - m_Z \cos \theta_W \sin \beta &0 \\
    & & 0 & -\mu_{\text{tot}} & - \frac{1}{\sqrt{2}} \lambda v \sin \beta \\
    & & & 0 & -\frac{1}{\sqrt{2}} \lambda v \cos \beta \\
    & & & & m_{\text{N}}
    \end{array}
    \right), \label{eq:mmn}
\end{equation}
where  $M_1$ and $M_2$ represent gaugino soft-breaking masses, and we define $s_W \equiv \sin \theta_W$, and $c_w \equiv \cos \theta_W$.
The physical mass eigenstates $\tilde{\chi}^0_i$ are obtained through diagonalization using a rotation matrix $N$: 
\begin{eqnarray}  \label{Mass-eigenstate-neutralino}
    \tilde{\chi}_i^0 = N_{i1} \psi^0_1 +   N_{i2} \psi^0_2 +   N_{i3} \psi^0_3 +   N_{i4} \psi^0_4 +   N_{i5} \psi^0_5,
\end{eqnarray}
where the index $i$ runs from 1 to 5, with states ordered by increasing mass. The coefficients $N_{i3}$ and $N_{i4}$ represent the $\tilde{H}_d^0$ and $\tilde{H}_u^0$ components in $\tilde{\chi}_i^0$, respectively. We call $\tilde{\chi}_1^0$ the Higgsino-dominated DM if $( N_{13}^2 + N_{14}^2 ) > 0.5$.

When the gauginos become very massive, they effectively decouple from the Higgsino-Singlino system. This decoupling reduces the original $5 \times 5$ neutralino mass matrix to a simpler $3 \times 3$ matrix that captures the essential Higgsino DM properties. Working in the basis 
($\tilde{H}_1 \equiv (\tilde{H}_d + \tilde{H}_u)/\sqrt{2}$, $\tilde{H}_2 \equiv (\tilde{H}_d - \tilde{H}_u)/\sqrt{2}$, $\tilde{S}$), this reduced matrix takes the form:
\begin{equation}
  M_{\tilde{\chi}^0} = \left(
  \begin{array}{ccc}
-\mu_{\rm tot}      &  0    & - \frac{1}{2} \lambda v (\sin \beta + \cos \beta)    \\
    &    \mu_{\rm tot}      & -\frac{1}{2} \lambda v (\sin \beta - \cos \beta)  \\
      &         &      m_N    
\end{array}
  \right). \label{eq:mmn}
\end{equation}
The corresponding mass eigenstates in this reduced basis can be expressed as:
\begin{eqnarray}  \label{Mass-eigenstate-neutralino-1}
    \tilde{\chi}_i^0 = N_{i1}^\prime \tilde{H}_1 +   N_{i2}^\prime \tilde{H}_2 +  N_{i3}^\prime \tilde{S},
\end{eqnarray}
where the elements of the new rotation matrix $N^\prime$ are given by $N_{i1}^\prime \equiv (N_{i3} + N_{i4})/\sqrt{2}$,   $N_{i2}^\prime \equiv (N_{i3} - N_{i4})/\sqrt{2}$, and $N_{i3}^\prime \equiv N_{i5}$. 

Our subsequent analysis focuses on scenarios featuring substantial mixing between $\tilde{H}_1$ and $\tilde{S}$ states. We parameterize  $m_N$ using a positive number $d$, such that $m_N = -(1 + d) \mu_{tot}$, and we work under the assumptions 
$\mu_{\rm tot} > 0$ and $\mu_{\rm tot} \gg d \mu_{\rm tot} > \lambda v$.  Following the methodology outlined in Ref.~\cite{Pierce:2013rda}, we derive approximate expressions for neutralino masses and mixing parameters:  
\begin{eqnarray}
m_{\tilde{\chi}_{1,3}^0} & \simeq & - \mu_{\rm tot} \pm \frac{\sqrt{(d \mu_{\rm tot})^2 + \lambda^2 v^2 (1 + \sin 2 \beta )} \mp d \mu_{\rm tot}}{2}, \nonumber \\
m_{\tilde{\chi}_2^0} & \simeq &  \mu_{\rm tot} + \frac{\lambda^2 v^2 (1 - \sin 2 \beta)}{8 \mu_{\rm tot}},  
\end{eqnarray} 
which implies 
\begin{eqnarray}
m_{\tilde{\chi}_{1}^0} \simeq - \mu_{\rm tot} + \frac{\lambda^2 v^2 (1 + \sin 2 \beta )}{4 d \mu_{\rm tot}}, \quad m_{\tilde{\chi}_{2}^0} \simeq \mu_{\rm tot}, \quad 
m_{\tilde{\chi}_{3}^0} \simeq - \mu_{\rm tot} - d \mu_{\rm tot},   \label{Neutralino-mass-approximation}
\end{eqnarray}
if $d \mu_{\rm tot}$ is significantly larger than $\lambda v$, and 
\begin{eqnarray} 
\frac{N_{i1}^\prime}{N_{i3}^\prime} =  - \frac{\lambda v (\sin \beta +\cos \beta)}{2 (\mu_{\rm tot} + m_{\tilde{\chi}_i^0})}, \quad \frac{N_{i2}^\prime}{N_{i3}^\prime} = \frac{\lambda v (\sin \beta - \cos \beta)}{2 (\mu_{\rm tot} - m_{\tilde{\chi}_i^0})}.   \label{Neutralino-mixing-1}
\end{eqnarray}
The Singlino component in $\tilde{\chi}_i^0$ can be expressed as
\begin{eqnarray}
    \left ( N_{i3}^{\prime} \right )^2 & = & \left(1+ \left ( \frac{ N^{\prime}_{i1}}{N^{\prime}_{i3}} \right)^2 + \left ( \frac{N^{\prime}_{i2}}{N^{\prime}_{i3}} \right )^2 \right)^{-1} \nonumber \\
    & = & \frac{2 (m_{\tilde{\chi}_i^0}^2  - \mu_{\rm tot}^2)^2}{2 (m_{\tilde{\chi}_i^0}^2  - \mu_{\rm tot}^2)^2 + \lambda^2 v^2 (m_{\tilde{\chi}_i^0}^2 + \mu_{\rm tot}^2 - 2 \sin 2 \beta  \mu_{\rm tot} m_{\tilde{\chi}_i^0}) }. 
\end{eqnarray}
For the parameter space relevant to our study, which is characterized by $\lambda \sim {\cal{O}}(0.1)$ and $\mu_{\rm tot} > 600~{\rm GeV}$, the rotation matrix $N^\prime$ can be approximated by a simpler form:
\begin{equation}
  N^\prime \simeq \left(
  \begin{array}{ccc}
 \cos \theta     &  0    & \sin \theta    \\
 0   &    1    &  0  \\
 -\sin \theta      &    0     & \cos \theta      
\end{array}
  \right),  \label{H-S-Mixing-1}
\end{equation}
where the Higgsino-Singlino mixing angle is given by
\begin{eqnarray}
\tan \theta & = & \frac{ d \mu_{\rm tot} - \sqrt{(d \mu_{\rm tot})^2 + \lambda^2 v^2 (1 + \sin 2 \beta )}}{\lambda v (\sin \beta +\cos \beta)} \nonumber \\
& \simeq & - \frac{\lambda v}{2 d \mu_{\rm tot}} (\sin \beta +\cos \beta),  \quad \quad {\rm if}\ d \mu_{\rm tot} \gg \lambda v.   \label{mixing angle}
\end{eqnarray} 
This simplified representation of $N^\prime$ significantly facilitates the analysis presented in subsequent sections.

\subsection{\label{RD}DM Relic Density}

The relic density of Weakly Interacting Massive Particle (WIMP) DM can be calculated from its thermally averaged annihilation cross-section in the non-relativistic limit. In the absence of co-annihilation, this cross-section is given by~\cite{1995Supersymmetric}:
\begin{equation}
\label{thermalcs}
\left\langle \sigma_Av \right\rangle = a + b\left\langle v^2 \right\rangle + \mathcal{O}(\left\langle v^4 \right\rangle) \approx a + 6\frac{b}{x},
\end{equation}
where $a$ represents the velocity-independent $s$-wave contribution, and $b$ incorporates both $s$- and $p$-wave components. The dimensionless parameter $x \equiv m_{\tilde{\chi}_1^0}/T$ characterizes the ratio of DM mass to the thermal bath temperature $T$ in the early Universe. 
Furthermore, by solving the Boltzmann equation from the freeze-out temperature to the present day, one obtains the current relic density~\cite{Baum:2017enm}:
\begin{eqnarray}
    \Omega h^2 = 0.12\left(\frac{80}{g_*}\right)^{1/2}\left(\frac{x_F}{25}\right) \left( \frac{2.3\times 10^{-26} \mathrm{cm^3/s}}{\langle \sigma v\rangle_{x_F}}\right),
\end{eqnarray}
where $g_*\sim 80$ denotes the effective number of relativistic degrees of freedom at freeze-out, and $x_F \equiv m/T_F \sim 25$ is determined by the freeze-out condition~\cite{Baum:2017enm}. In scenarios with co-annihilation, the relic density is computed in a similar way, as shown by Eq.~15 in Ref.~\cite{Griest:1990kh}.

For the Higgsino DM in this study, the observed relic density is primarily achieved through two types of processes: direct annihilation $\tilde{\chi}_1^0 \tilde{\chi}_1^0 \to X Y$ (where $X$ and $Y$ are SM particles) and co-annihilation processes involving channels such as $\tilde{\chi}_{1,2,3}^0 \tilde{\chi}_1^\pm$, $\tilde{\chi}_{1}^0 \tilde{\chi}_{2}^0$, $\tilde{\chi}_1^\pm \tilde{\chi}_1^\mp \to X Y$.
The effective annihilation cross-section is given by~\cite{Griest:1990kh,Baker:2015qna}:
\begin{eqnarray}  
\sigma_{eff} = \sum_{ij} \sigma (\tilde{\chi}_i \tilde{\chi}_j \to X Y) \times \frac{g_i g_j}{g_{eff}^2}(1 + \Delta_i)^{3/2}(1 + \Delta_j)^{3/2}\exp[-x(\Delta_i + \Delta_j)  \label{eff-cross-section}
\end{eqnarray}
where $\Delta_i \equiv (m_{\tilde{\chi}_i} - m_{\tilde{\chi}_1^0})/m_{\tilde{\chi}_1^0}$, $g_i$ denotes the degrees of freedom for $\tilde{\chi}_i$, and  
\begin{eqnarray}  
g_{eff} \equiv \sum_{i=1}^{N} g_i(1 + \Delta_i)^{3/2}\exp(-x\Delta_i). 
\end{eqnarray}  
For the parameter space delineated by $\lambda \sim {\cal{O}}(0.1)$ and $\mu_{\rm tot} > 600~{\rm GeV}$, the GNMSSM predictions on the cross-sections $\sigma (\tilde{\chi}_i \tilde{\chi}_j \to X Y)$ can be approximated by scaling the corresponding MSSM ones with factors $f_{\tilde{\chi}_i \tilde{\chi}_j}$, which are $f_{\tilde{\chi}_1^0 \tilde{\chi}_1^0 } = \cos^4 \theta $,  $f_{\tilde{\chi}_1^0 \tilde{\chi}_1^\pm } = \cos^2 \theta $, $f_{\tilde{\chi}_3^0 \tilde{\chi}_1^\pm } = \sin^2 \theta $, 
$f_{\tilde{\chi}_1^0 \tilde{\chi}_2^0 } = \cos^2 \theta \sin^2 \theta $, and $f_{\tilde{\chi}_2^0 \tilde{\chi}_1^\pm } = f_{\tilde{\chi}_1^\pm \tilde{\chi}_1^\mp } = 1 $. 
This relationship is verified by calculating the $s$-wave contributions to the cross-sections for $\tilde{\chi}_1^0 \tilde{\chi}_1^0 \to W W$ and $\tilde{\chi}_1^0 \tilde{\chi}_1^0 \to Z Z$, which proceed via $t$-channel exchanges of $\tilde{\chi}_1^\pm$ and  $\tilde{\chi}_{1,2,3}^0$, respectively:  
\begin{eqnarray}
\langle \sigma v \rangle_{WW}^s \simeq \frac{1}{128 \pi m_{\tilde{\chi}_1^0}^2} g^4 \cos^4{\theta}, \quad \langle \sigma v \rangle_{ZZ}^s \simeq  \frac{1}{256 \pi m_{\tilde{\chi}_1^0}^2} \frac{g^4}{c_W^4} \cos^4{\theta}.  
\end{eqnarray}  
These results differ from the MSSM predictions~\cite{1995Supersymmetric,Drees:1992am} by a factor $\cos^4 \theta$. It arises from two sources: a $\cos \theta$ scaling of the $\tilde{\chi}_1^0 \tilde{\chi}_1^\pm W $ and 
$\tilde{\chi}_1^0 \tilde{\chi}_2^0 Z $ couplings, and the strong suppression of $\tilde{\chi}_1^0 \tilde{\chi}_1^0 Z $ and  $\tilde{\chi}_1^0 \tilde{\chi}_3^0 Z $ couplings in the GNMSSM.  

Overall, these findings demonstrate that both the mass splitting among Higgsino-like particles and the Higgsino-Singlino mixing are crucial in determining the final relic density.

\subsection{\label{DMDD}DM Direct Detection}

In scenarios where squarks are extremely massive, the SI scattering of DM off nucleons predominantly arises from the $t$-channel exchange of $CP$-even Higgs bosons. The cross-section is given by~\cite{1995Supersymmetric,Drees:1992rr,Drees:1993bu}
\begin{eqnarray}
   \sigma^{\rm SI}_{N} = \frac{4 \mu_r^2}{\pi} |f^{N}|^2, \quad
	f^{N} =  \sum_{i}^3 f^{N}_{h_i} = \sum_{i}^3 \frac{C_{ \tilde {\chi}^0_1 \tilde {\chi}^0_1 h_i} C_{N N h_i }}{2m^2_{h_i} }. \label{SI-cross}
\end{eqnarray}
Here, $N=p, n$ denotes a proton ($p$) or neutron ($n$), and $\mu_r \equiv m_N m_{\tilde {\chi}^0_1} /(m_N+m_{\tilde {\chi}^0_1})$ represents the reduced mass of the DM-nucleon system. The coupling $C_{NN h_i}$ characterizes the strength 
of Higgs-nucleon interaction and is written as 
\begin{eqnarray}
C_{NN h_i} = -\frac{m_N}{v}
\left[ F^{N}_d \left( V_{h_i}^{\rm SM}- \tan\beta V_{h_i}^{\rm NSM}\right)+F^{N}_u \left(V_{h_i}^{\rm SM}+\frac{1}{\tan\beta} V_{h_i}^{\rm NSM} \right)
\right]. \quad \label{C-hNN_S}
\end{eqnarray}
In this expression, $F^N_d$ and $F^N_u$ are defined by $F^{N}_d \equiv f^{(N)}_d+f^{(N)}_s+\frac{2}{27}f^{(N)}_G$ and $F^{N}_u \equiv f^{(N)}_u+\frac{4}{27}f^{(N)}_G$, where
the nucleon form factors $f^{(N)}_q \equiv m_N^{-1}\left<N|m_qq\bar{q}|N\right>$ for $q=u,d,s$ represent the normalized contribution of light quarks to the nucleon mass. The term $f^{(N)}_G \equiv 1-\sum_{q=u,d,s}f^{(N)}_q$ accounts for the heavy quarks contributions (see, e.g, Ref.~\cite{Drees:1993bu,Drees:1992rr}). Under the default settings of the micrOMEGAs package for $f_q^{(N)}$~\cite{Belanger:2008sj,Alarcon:2011zs,Alarcon:2012nr}, one obtains $F_u^{p} \simeq F_u^n \simeq 0.15$, $F_d^{p} \simeq F_d^n \simeq 0.13$, implying that $\sigma^{\rm SI}_{p}$ is approximately equal to $\sigma^{\rm SI}_{n}$ without strong cancellations between different contributions. This approximate degeneracy, however, is usually broken for small $\sigma^{\rm SI}_{p}$ and $\sigma^{\rm SI}_{n}$ as suggested by recent LZ results~\cite{LZ:2024zvo}.  

Regarding the Higgs coupling to the DM pair, $C_{\tilde{\chi}^0_1 \tilde{\chi}^0_1 h_i }$, it is approximated by~\cite{Ellwanger:2009dp}
\begin{eqnarray}
C_{\tilde{\chi}^0_1 \tilde{\chi}^0_1 h_i }   & \simeq & 
\lambda \sin \theta \cos \theta \big[ V_{h_i}^{\rm NSM}  ( \cos \beta - \sin \beta ) + V_{h_i}^{\rm SM} (\cos \beta + \sin \beta) \big] \nonumber \\
&-& \sqrt{2} \, \kappa \, V_{h_i}^S \sin^2 \theta + \frac{\lambda}{\sqrt{2}} V_{h_i}^S \cos^2 \theta,
\end{eqnarray}
where the Higgsino-Singlino mixing angle is defined in Eq.~(\ref{mixing angle}).

In the limit where the charged Higgs bosons are very massive, one finds $V_h^{\rm NSM} \sim 0$, and $V_{h_s}^{\rm NSM} \sim 0$. Furthermore, as discussed in Ref.~\cite{Meng:2024lmi}, the $H$-mediated contribution can be safely neglected due to its suppression by a factor $1/m_H^4$. Consequently, assuming $F_u^N = F_d^N$, the SI cross-section can be simplified to~\cite{Cao:2021ljw,Zhou:2021pit}
\begin{eqnarray}
\sigma^{\rm SI}_{N} & \simeq & 5 \times 10^{-45}\,{\rm cm^2}~\left(\frac{\cal{A}}{0.1}\right)^2, \label{SI}
\end{eqnarray}
with
\begin{eqnarray}
{\cal{A}} &=& \left( \frac{125 \rm GeV}{m_{h}}\right)^2 V_h^{\rm SM} C_{ \tilde {\chi}^0_1 \tilde {\chi}^0_1 h} +  \left( \frac{125 \rm GeV}{m_{h_s}} \right)^2  V_{h_s}^{\rm SM} C_{ \tilde {\chi}^0_1 \tilde {\chi}^0_1 h_{s}} \nonumber \\
  & = & \lambda (\cos \beta + \sin \beta) \sin\theta \cos \theta \left[ \left(\frac{125\,\text{GeV}}{m_h}\right)^2 \left(V_h^{\rm SM}\right)^2 + \left(\frac{125\,\text{GeV}}{m_{h_s}}\right)^2 \left(V_{h_s}^{\rm SM}\right)^2 \right] \nonumber \\
 & & + \frac{\lambda}{\sqrt{2}}\cos^2\theta \left[ \left(\frac{125\,\text{GeV}}{m_h}\right)^2 V_h^{\rm SM} V_h^S + \left(\frac{125\,\text{GeV}}{m_{h_s}}\right)^2 V_{h_s}^{\rm SM} V_{h_s}^S \right] \nonumber \\
  & & - \sqrt{2} \kappa \sin^2 \theta \left[ \left(\frac{125\,\text{GeV}}{m_h}\right)^2 V_h^{\rm SM} V_{h}^S + \left(\frac{125\,\text{GeV}}{m_{h_s}}\right)^2 V_{h_s}^{\rm SM} V_{h_s}^S \right]. \label{SIA}
\end{eqnarray}
Given that $V_h^{\rm SM} \simeq V_{h_s}^{\rm S} \simeq 1$, $V_h^{\rm S} \simeq - V_{h_s}^{\rm SM}$, $\cos \theta \sim 1$, and that $V_h^{\rm S}, V_{h_s}^{\rm SM}, \sin \theta$ are proportional to $\lambda$, we have the following observations: 
\begin{itemize}
\item The leading contributions in the first, second, and third terms of ${\cal{A}}$ scale as $\lambda^2$, $\lambda^2$, and $\kappa \lambda^3$, respectively. 
\item Cancellations occur among the contributions from different Higgs bosons, with $h_s$ potentially exerting a significant influence when it is light. 
\item A sufficiently large $\kappa$ with an appropriate sign can counterbalance the other contributions.
\item The sign of $V_h^{\rm S}$, as given in Eq.~(\ref{Approximations}), can induce cancellation between the first and second terms of  ${\cal{A}}$.  
\end{itemize}

On the other hand, the SD scattering cross section is primarily due to $t$-channel $Z$ boson exchange. It can be expressed as~\cite{Chalons:2012xf}
\begin{eqnarray}
\sigma_{N}^{\rm SD} = C_N \times \left(\frac{C_{\tilde{\chi}_1^0 \tilde{\chi}_1^0 Z}}{0.1}\right)^2, \label{SD}
\end{eqnarray}
where $C_n = 3.1 \times 10^{-40} \, \text{pb}$  and $C_p = 4.0 \times 10^{-40} \, \text{pb}$ reflect differences in nuclear structure. The normalized coupling of DM to Z boson, $C_{\tilde{\chi}_1^0 \tilde{\chi}_1^0 Z}$, is given by $C_{\tilde{\chi}_1^0 \tilde{\chi}_1^0 Z} \equiv N_{13}^2-N_{14}^2$. Notably, this coupling vanishes under the mixing angle approximation in Eq.~(\ref{H-S-Mixing-1}), reflecting a strong suppression of the SD cross-section. However, using the exact expressions for the neutralino mixing matrix elements given in Eq.~(\ref{Neutralino-mixing-1}), one obtains 
\begin{eqnarray}
  C_{\tilde{\chi}_1^0 \tilde{\chi}_1^0 Z}  \approx \frac{\mu_{\rm tot} + m_{\tilde{\chi}_1^0}}{\mu_{\rm tot}} \frac{\lambda^2 v^2 \cos 2 \beta}{\lambda^2 v^2 (1 + \sin 2 \beta) + 4 (\mu_{\rm tot} + m_{\tilde{\chi}_1^0})^2}. \label{SD-approximation}
\end{eqnarray}
Using the approximation of $m_{\tilde{\chi}_1^0}$ in Eq.~(\ref{Neutralino-mass-approximation}), one can infer that  $C_{\tilde{\chi}_1^0 \tilde{\chi}_1^0 Z}$ is suppressed by a factor of $\lambda^2 v^2/(d \mu_{\rm tot}^2)$.  It vanishes when $\tan \beta = 1$, which was called the blind-spot for the SD scattering in literatures. 

\section{\label{numerical study}Numerical Study}

This section outlines our sampling methodology and examines the distinctive features of the Higgsino DM in the GNMSSM. Our numerical analysis follows a systematic approach using state-of-the-art computational tools.
We begin by implementing the GNMSSM model routines using \textsf{SARAH-4.15.3}~\cite{Staub:2008uz,Staub:2012pb,Staub:2013tta,Staub:2015kfa}. The particle spectrum and low-energy flavor observables are then computed using SPheno-4.0.5~\cite{Porod_2012,Porod:2003um,Belanger:2014hqa} and FlavorKit~\cite{Porod:2014xia}, respectively. For DM physics observables, we employ \textsf{MicrOMEGAs-5.0.4}~\cite{Belanger:2001fz, Belanger:2005kh,Belanger:2006is,belanger2010micromegas,Belanger:2010pz,Belanger:2013oya,Barducci:2016pcb,Belanger:2018mqt}. The parameter space exploration is conducted using a parallelized version of the MultiNest algorithm~\cite{Feroz:2008xx} within a modified version of \textsf{EasyScan}~\cite{Shang:2023gfy}. This approach efficiently identifies high-likelihood regions, detects multiple modes, and provides robust Bayesian evidence estimates. To identify viable parameter regions and extract physical insights, we analyze the obtained samples using profile likelihood (PL) methods within the Frequentist statistical framework~\cite{Fowlie:2016hew}. In our analysis, we also integrate the latest  $125~{\rm GeV}$ Higgs data and findings from the LZ dark matter direct detection experiment to investigate their impact on the model’s characteristics.

\begin{table}[tbp]
  \centering  
  \vspace{0.3cm}  
  \resizebox{0.7\textwidth}{!}{  
  \begin{tabular}{c|c|c|c|c|c}  
  \hline  
  Parameter & Prior & Range & Parameter & Prior & Range   \\
  \hline  
  $\kappa$  & Flat & $-0.7$--$0.7$ & $\tan \beta$ & Flat & $5$--$30$ \\   
  $\lambda$ & Flat & $0$--$0.5$    & $d^{\prime}$ & Flat & $0.1$--$0.7$ \\
  $\delta$ & Flat & $-0.2$--$0.2$ & $m_B/{\rm TeV}$ & Flat & $0.05$--$0.3$ \\
  $A_t/{\rm TeV}$ & Flat & $2.5$--$5$ & $\mu_{\rm tot}/{\rm TeV}$ & Flat & $0.6 $--$ 1.1$ \\
  \hline  
  \end{tabular}} 
  \caption{Parameter space explored in this study, where $d^{\prime}$ is defined as $d^\prime \equiv d/0.1$ for simplicity. All input parameters are assigned flat distributions in their priors due to their well-defined physical interpretations.  To concentrate on the effects of Higgsino-Singlino mixing on the DM properties and also to evade strong constraints from the LZ experiment, we set the gaugino masses $M_1$ and $M_2$ at $2~{\rm TeV}$ to ensure that the Higgsino-Gaugino mixing remains small. Additionally, since the soft trilinear coefficients for the third-generation squarks, 
$A_t$ and $A_b$, can significantly influence the SM-like Higgs boson mass through radiative corrections, we establish $A_t = A_b$ and allow them to vary.
All other dimensional SUSY parameters not central to the analysis are kept fixed: $\xi^\prime = 0$, $m_C = 400~{\rm GeV}$, $v_s = 600~{\rm GeV}$, 
and a universal value of $3~{\rm TeV}$ for the remaining parameters, consistent with constraints from the LHC new physics searches. 
All parameters are defined at the renormalization scale  $Q_{input} = 1~{\rm TeV}$. 
The final parameter space is derived through multiple scans over significantly wider ranges.}
  \label{ScanRange}   
\end{table}  

\subsection{\label{scan}Research Strategy}

Through a process of trial and error, we identify the GNMSSM parameter space to be explored, as shown in Table~\ref{ScanRange}. Given the high degree of fine-tuning in the Higgsino DM region, we partition the $\lambda-\mu_{\rm tot}$
plane into a  $10 \times 10$ grid to capture the underlying subtle physics more accurately. Each grid subsection  undergoes a parallelized MultiNest scan with 6000 live points, configured by setting the parameter of the algorithm, $nlive$, at 6000.

The rationale for each parameter range is grounded in theoretical consistency, phenomenological viability, and the specific objectives of our analysis, as detailed below:

\begin{itemize}
    \item \textbf{Dimensionless Couplings ($\boldsymbol{\lambda, \kappa}$):} The dimensionless couplings $\lambda$ and $\kappa$ are fundamental to the model's structure. Their ranges are primarily constrained to ensure the validity of the theory up to the Grand Unification(GUT) scale~\cite{Ellwanger:2009dp}, with additional considerations as follows:
    \begin{itemize}
        \item We set $\lambda \in [0, 0.5]$ to ensure substantial mixing between the Higgsino and singlino, which is central to the neutralino phenomenology we study. Larger values of $\lambda$s become disfavored by the Higgs data fit and the DM-nucleon scattering constraints discussed below. 
        \item The range $\kappa \in [-0.7, 0.7]$ allows for a significant variation in the singlet self-coupling while maintaining theoretical consistency. 
    \end{itemize}
    
    \item \textbf{Higgs Sector and Radiative Corrections ($\boldsymbol{A_t, \tan\beta}$):} These parameters are crucial for obtaining a realistic Higgs sector:
    \begin{itemize}
        \item The trilinear soft-breaking parameter $A_t$ is scanned within [2.5, 5] TeV. This range is motivated by the requirement to generate sufficient radiative corrections to achieve a Higgs boson mass of 125 GeV. 
        \item We set $\tan \beta \in [5, 30]$. The lower bound is chosen to ensure a substantial tree-level Higgs boson mass and meanwhile remaining consistent with LEP constraints on the Higgs sector~\cite{Mahmoudi:2010xp}. The upper bound helps  avoid large flavor-violating effects, particularly from Br($b \to s \gamma$) constraints, and ensures the stability of the electroweak vacuum against color breaking minima~\cite{King:2012is}.
    \end{itemize}
        
    \item \textbf{Neutralino Mass and Mixing Phenomenology ($\boldsymbol{\mu_{\rm tot}, d', \delta}$):} This set of parameters directly controls the DM candidate's properties:
    \begin{itemize}
        \item The effective Higgsino mass parameter $\mu_{\rm tot}$ is varied in the range [0.6, 1.1] TeV. This allows us to investigate a phenomenologically interesting region where the Higgsino-like charginos and neutralinos have masses from approximately 600 GeV to 1 TeV. This range is explicitly chosen to test compatibility with the observed DM relic density and stringent direct detection limits.
        \item The parameter $d^\prime $ (defined as $d' \equiv d / 0.1$) parameterizes the mass splitting between the singlino and Higgsino states via the relation $m_N = -(1 + d) \mu_{\rm tot}$. We scan $d^\prime  \in [-0.2, 0.2]$ to explore scenarios with small to moderate mass splittings, which have important consequences for Higgsino-Singlino coannihilation and relic density calculations.
        \item The parameter $\delta$ is defined via $M_{S_{23}}^2 =\sqrt{2} \lambda \delta v \mu_{tot}$  and serves as a measure of cancellation in the CP-even Higgs mixing matrix element  $M^2_{S,23}$ (see text below Eq. 2.14 in our manuscript). By scanning $\delta \in [-0.2, 0.2]$, we allow for moderate cancellations—a crucial mechanism that enables moderately large $\lambda$ values to remain consistent with constraints from Higgs data fits~\cite{Lian:2024smg}.
     \end{itemize}
     \item \textbf{Singlet CP-even Higgs boson mass $m_B$:} This parameter may play a significant role in DM-nucleon scattering, where smaller values usually lead to more substantial impact on the scattering cross-section. Its lower bound is motivated by the need to avoid excessively large branching ratios for the decay channel $h \to h_s h_s$, which would conflict with current Higgs decay measurements~\cite{Heng:2025wng}. Its upper bound is chosen to ensure that the effects of the singlet Higgs remain phenomenologically significant and observable, as verified by our numerical analysis.
\end{itemize}

During these scans, we construct a joint likelihood function that incorporates DM relic density, DM direct detection limits, and a range of other experimental constraints, including LHC Higgs data. For the relic density likelihood 
$\mathcal{L}_{\Omega h^2}$, we adopt a Gaussian distribution centered on the Planck collaboration's measurement~\cite{Planck:2018vyg}, with a $10\%$ theoretical uncertainty to account for systematic calculation errors. For direct detection constraints from the LZ results in 2022~\cite{LZ:2022lsv}, we define  $\mathcal{L}_{\rm LZ}$ using a Gaussian distribution centered at zero with variance $\delta_\sigma^2 = ({\rm UL}_\sigma / 1.64)^2 + (0.2\sigma)^2$, 
where ${\rm UL}_\sigma$ represents the $90\%$ confidence-level upper limit on the SI scattering cross-section, and the $0.2\sigma$ term accounts for nuclear response and other theoretical uncertainties~\cite{Matsumoto:2016hbs}. Furthermore, since the SI cross-section for DM-proton scattering, $\sigma_{p}^{\text{SI}}$, may significantly deviate from that for DM-neutron scattering, $\sigma_{n}^{\text{SI}}$, when the scattering rates are tiny, we define the effective SI cross-section as $\sigma_{\text{eff}}^{\text{SI}} = 0.169\sigma_{p}^{\text{SI}} + 0.347\sigma_{n}^{\text{SI}} + 0.484\sqrt{\sigma_{p}^{\text{SI}}\sigma_{n}^{\text{SI}}}$~\cite{Cao:2019aam}, which averages the abundance of different xenon isotopes in nature~\cite{XENON:2018voc}, and compare it with the published results of the LZ experiment.

The complete likelihood function is formulated as
\begin{eqnarray}  
  \mathcal{L} & \equiv & \mathcal{L}_{\Omega h^2} \times \mathcal{L}_{\rm LZ} \times \mathcal{L}_{\rm Const},  \nonumber \\
  \mathcal{L}_{\Omega h^2} &=& {\rm Exp}\left[ -\frac{1}{2}\left( \frac{\Omega h^2-0.120}{0.012 } \right)^2 \right], \quad  
  \mathcal{L}_{\rm LZ}= {\rm Exp} \left[-\frac{\sigma^{\rm SI}_{\rm eff}}{2\delta_{\sigma}^2} \right], \nonumber \\
  \mathcal{L}_{\rm Const} &=& \left\{ \begin{aligned} & 1, & &{\rm  Satisfying\ all\ experimental\ constraints} \\ & {\rm Exp}\left[-100\right]. & &{\rm Otherwise} \end{aligned} \right.   \label{Likelihood}  
\end{eqnarray}
Constraints included in $\mathcal{L}_{\text{Const}}$
encompass:
\begin{itemize}
\item \textbf{DM component:} $\tilde{\chi}_1^0$ should be Higgsino-dominated, i.e., $ N_{13}^2 + N_{14}^2 > 0.5$.  
\item \textbf{Higgs data fit:} The properties of the Higgs boson $h$ observed at the LHC must align with ATLAS and CMS measurements at $95\%$ confidence level. Assuming a $3~{\rm GeV}$ combined theoretical and experimental uncertainty for $m_h$, we computed the p-value of the fit using \textsf{HiggsSignals-2.6.2}~\cite{HS2013xfa,HSConstraining2013hwa,HS2014ewa,HS2020uwn} and required it larger than 0.05. 
\item \textbf{Extra Higgs searches:} Comprehensive searches for additional Higgs bosons at LEP, Tevatron, and LHC are implemented using \textsf{HiggsBounds-5.10.2}~\cite{HB2008jh,HB2011sb,HBHS2012lvg,HB2013wla,HB2020pkv}. 
\item \textbf{Indirect DM searches:} The Fermi-LAT collaboration has made years of observations of dwarf galaxies, limiting the annihilation cross section as a function of the DM mass. We employed the likelihood function proposed in Ref.~\cite{Carpenter:2016thc,Huang:2016tfo} to implement this constraint. We also used the latest MDHAT package, which incorporated 14 years of publicly available Fermi-LAT data from a set of 54 dwarf spheroidal galaxies, to improve this constraint~\cite{Boddy:2019kuw,Boddy:2024tiu}. 
\item \textbf{$B$-physics observables:} Branching ratios for $B_s \to \mu^+ \mu^-$ and $B \to X_s \gamma$ must conform to experimental measurements within $2\sigma$~\cite{pdg2020}. 
\end{itemize}
Notably, we exclude LHC SUSY search constraints as they do not impact our parameter space due to the high mass of the Higgsino DM under consideration.

\begin{figure*}[t]
  \centering
  \resizebox{1.0 \textwidth}{!}{
    \includegraphics{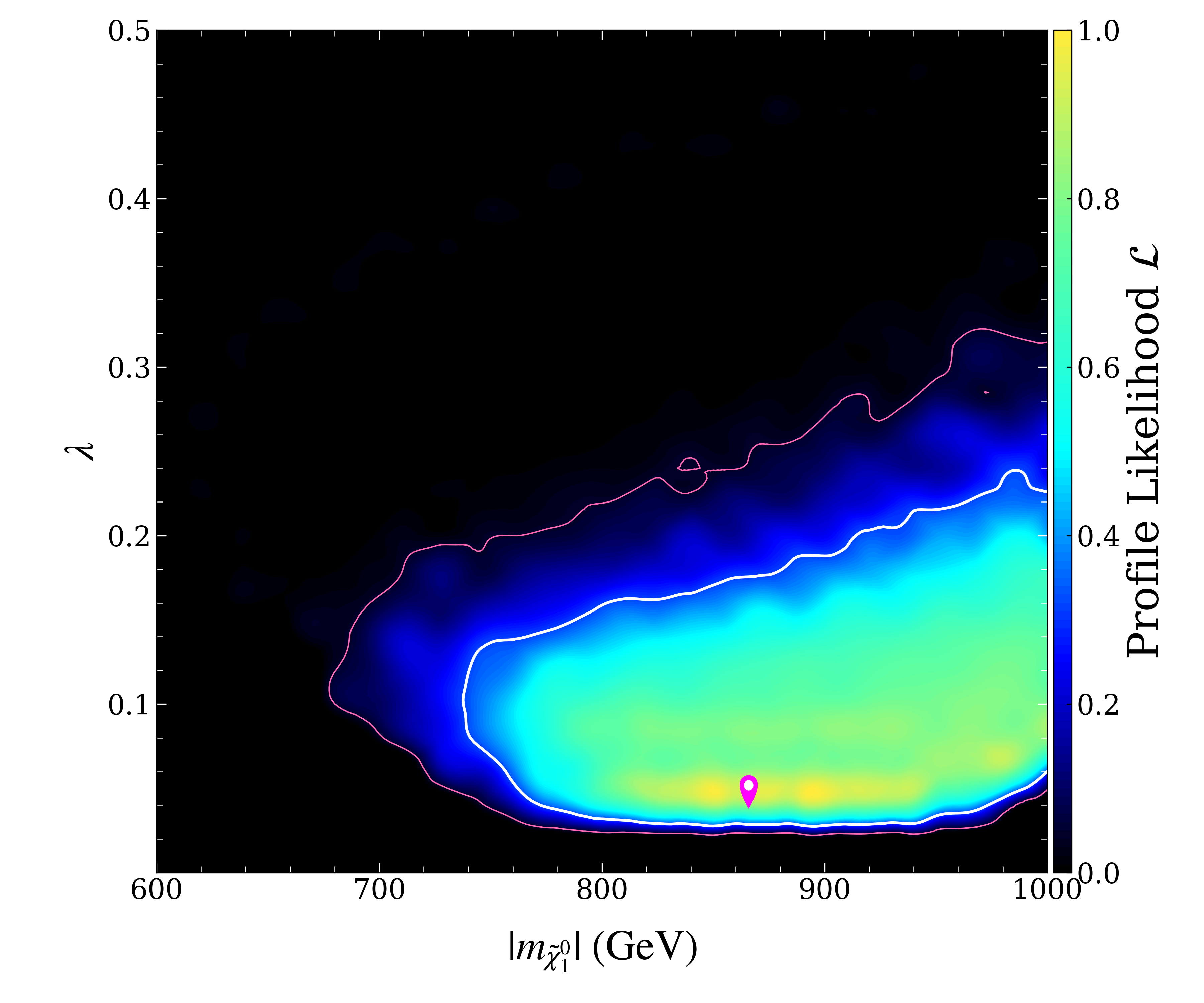}
    \includegraphics{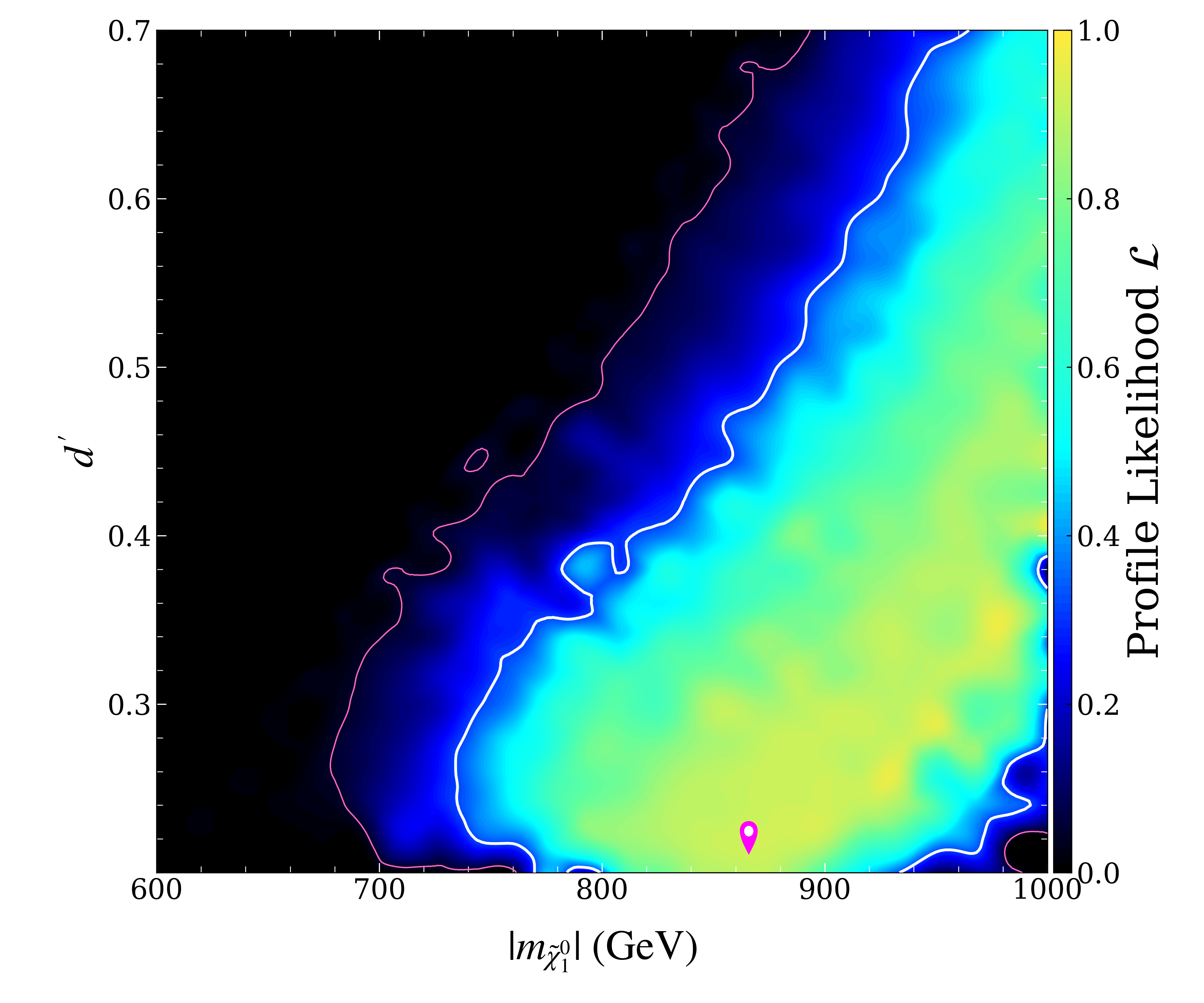}
  
  }
  
  \resizebox{1.0 \textwidth}{!}{
    \includegraphics{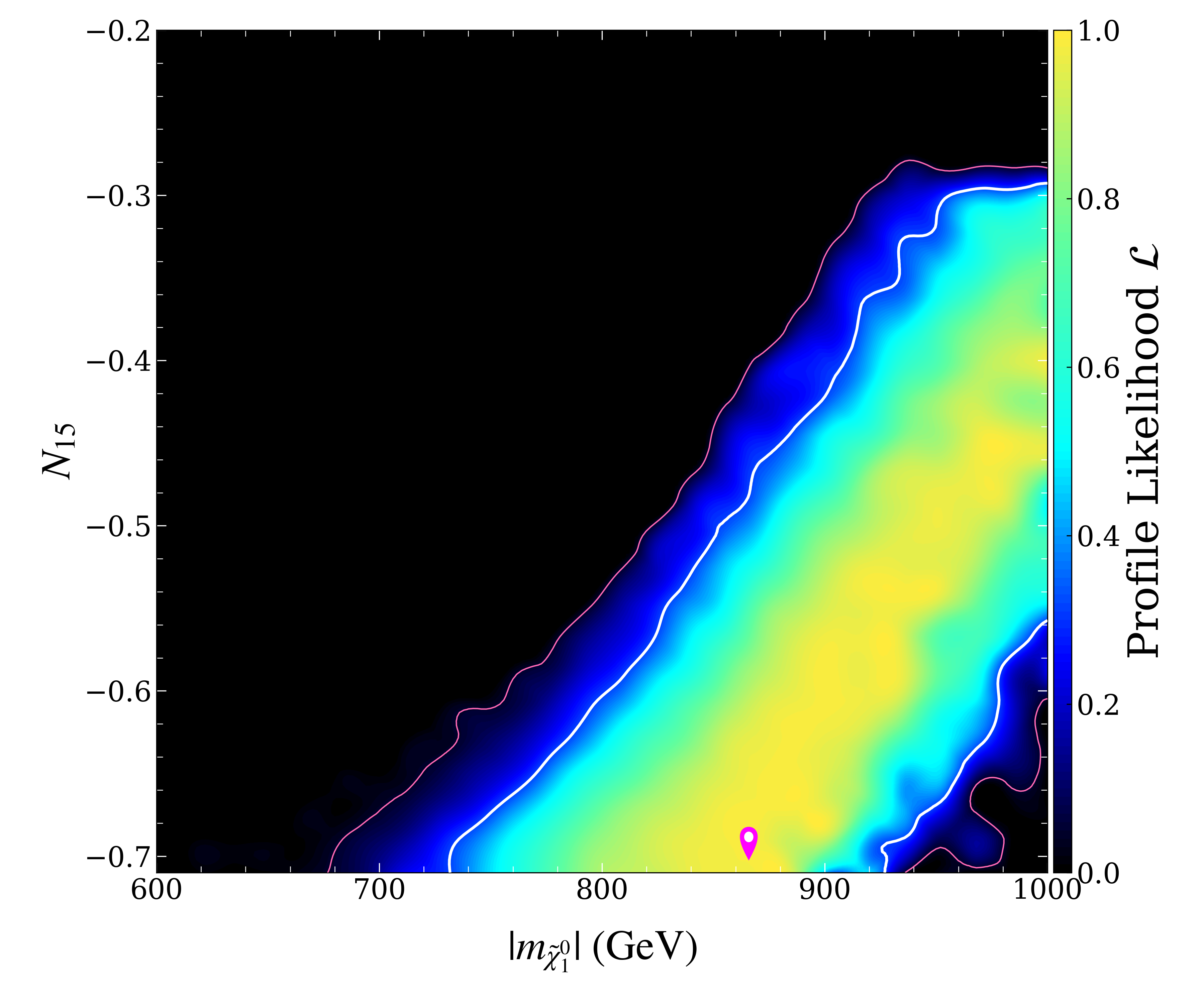}
    \includegraphics{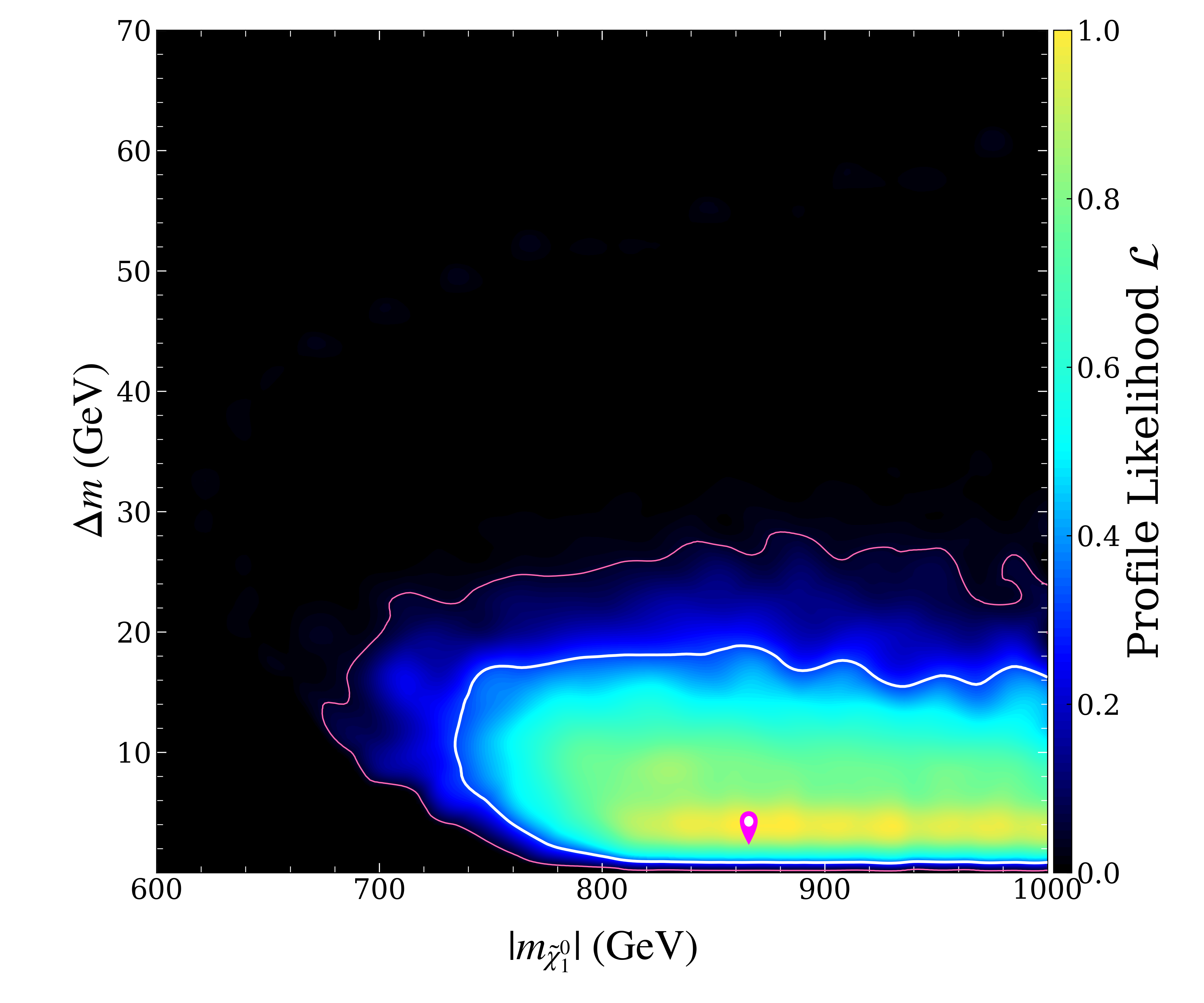}
  }
  \vspace{-0.0cm}
  \caption{Two-dimensional profile likelihoods of the function $\cal{L}$ in Eq.~(\ref{Likelihood}), projected
 onto the $|m_{\tilde{\chi}_1^0}|-\lambda$,  $|m_{\tilde{\chi}_1^0}|-d^\prime$,  $|m_{\tilde{\chi}_1^0}|-N_{15}$, and  $|m_{\tilde{\chi}_1^0}|-\Delta m$ planes, respectively. Here, $N_{15} \simeq \sin \theta$ represents the fraction of Singlino component in the Higgsino-dominated DM, and  $\Delta m \equiv m_{\tilde{\chi}_1^{\pm}}- |m_{\tilde{\chi}_1^0}|$ denotes the mass splittings between $\tilde{\chi}_1^\pm$ and $\tilde{\chi}_1^0$.   
 Since the best point (marked with pin symbol) yields  $\chi^2 \simeq 0$, the boundaries for $1 \sigma$ and $2 \sigma$ confidence intervals correspond to $\chi^2 \simeq 2.3$ and $\chi^2 \simeq 6.18$, respectively, indicated by white and red solid lines. This figure illustrates how the DM relic abundance constrains the parameter space of the GNMSSM.  In generating this figure, we have applied constraints from the 2022 LZ results~\cite{LZ:2022lsv} and the latest Higgs data fitting procedures described in Ref.~\cite{LZ:2022lsv}, as detailed in the preceding text. \label{Fig1}}
\end{figure*}

\subsection{Numerical Results}

Through our scanning procedure, we obtained a total of 2,146,266 parameter points. We then applied Higgs data fitting constraints to these samples using the latest version of HiggsTools-1.2~\cite{Bahl:2022igd}, and reanalyzed the previously scanned parameter points. This updated version of the HiggsSignals and HiggsBounds represents a significant advancement over HiggsSignals-2.6.2 by incorporating 22 additional physical observables, thereby raising the total number of observables for fitting Higgs properties to 129~\cite{Bahl:2022igd}. 

In our refitting process, we assumed an uncertainty of $3~{\rm GeV}$ for $m_h$ and required the samples to satisfy   $\chi^2_{\rm Higgs} - \chi^2_{\rm SM, 125} \lesssim 6.18$, where $\chi^2_{\rm Higgs}$ is the value obtained from the $125~{\rm GeV}$ Higgs data fit with \texttt{HiggsTools-1.2}, and $\chi^2_{\rm SM, 125} \simeq 176.3$ corresponds to a pure SM Higgs boson in the GNMSSM~\cite{Cao:2024axg}. After applying these constraints, we obtained 276,077 points, representing a reduction of approximately $87.1\%$.

We verified that the excluded samples are characterized by substantial values of $V_h^S$, which typically correspond to moderately large $\lambda$ values. The underlying reason for this behavior is that sizable $\lambda$ values are preferred in this study to achieve significant Higgsino-singlino mixing, but such values tend to create tension with the Higgs data constraints. 
 
Based on these points, we plotted the PL distributions across various two-dimensional planes\footnote{In frequentist statistics, the two-dimensional PL refers to the maximum likelihood value in a specific parameter space~\cite{Fowlie:2016hew}. For a given set of input parameters $\Theta \equiv\left(\Theta_1, \Theta_2, \cdots\right)$, this PL is obtained by maximizing the likelihood function while varying the other parameters:
\begin{equation}
\mathcal{L}\left(\Theta_A, \Theta_B\right)=\max _{\Theta_1, \cdots, \Theta_{A-1}, \Theta_{A+1}, \cdots, \Theta_{B-1}, \Theta_{B+1}, \cdots } \mathcal{L}(\Theta),
\end{equation}
At a given point $(\Theta_A, \Theta_B)$, $\mathcal{L}\left(\Theta_A, \Theta_B \right)$ reflects its capability to explain the experimental data, or alternatively, the data's preference for the parameter space of the theory. Correspondingly, one may define $\chi^2$ for the PL function as  $\chi^2 \left(\Theta_A,\Theta_B \right) \equiv - 2 \ln \mathcal{L}\left(\Theta_A, \Theta_B\right)$ to measure the point's capability in explaining the data.}, enabling us to analyze the characteristics of Higgsino DM and its underlying physical mechanisms.

We first examine how the DM relic abundance depends on the model parameters. Analysis of the scanned parameter points reveals that in most cases, DM achieves the experimentally measured relic abundance primarily through co-annihilation processes $\tilde{\chi}_1^0 \tilde{\chi}_1^\pm, \tilde{\chi}_2^0 \tilde{\chi}_1^\pm \to X Y $, where $X$ and $Y$ denote SM particles, along with the annihilation process $\tilde{\chi}_1^0 \tilde{\chi}_1^0 \to Z Z, W W, t \bar{t}$. According to the relevant equations in section~\ref{RD} for very massive gaugino cases, the relic density primarily depends on the mixing angle $\theta$ with $\sin \theta \simeq N_{15}$ and mass splitting $\Delta m \equiv m_{\tilde{\chi}_1^\pm} – |m_{\tilde{\chi}_1^0}|$, which are ultimately determined by parameters $\lambda$, $\mu_{\rm tot} \simeq |m_{\tilde{\chi}_1^0}|$, and $d^\prime$. Fig.~\ref{Fig1} shows the PL distribution of the parameter points projected onto the $|m_{\tilde{\chi}_1^0}|-\lambda$, $|m_{\tilde{\chi}_1^0}|-d^\prime$,  $|m_{\tilde{\chi}_1^0}|-N_{15}$, and $|m_{\tilde{\chi}_1^0}|-\Delta m$ planes, where white and red solid lines represent the boundaries of $1\sigma$ and $2 \sigma$ confidence intervals, respectively. Given that the best-fit point has $\chi^2 \simeq 0$, these contour lines correspond to $\chi^2 \simeq 2.3$ and $\chi^2 \simeq 6.18$, respectively. From Fig.~\ref{Fig1}, we can draw the following conclusions: 
\begin{itemize}
\item In the Higgsino DM scenario of the GNMSSM, due to mixing between Higgsino and Singlino, DM with mass as low as $670~{\rm GeV}$ can yield a relic density consistent with the experimental measurements. In contrast, the Higgsino DM in the MSSM requires a mass of approximately $1.1~{\rm TeV}$ to obtain an appropriate relic density, with a typical $\Delta m$ value of less than $1~{\rm GeV}$ after considering the constraints from latest LZ results (see point P2 in Table~\ref{Tab2}). This indicates that GNMSSM has a significantly larger parameter space for producing viable Higgsino DM than the MSSM.
	
\item When DM mass is around $670~{\rm GeV}$, relatively large values of $|N_{15}|$ and $\Delta m$ (specifically $N_{15} \simeq -0.70$, $\Delta m \simeq 15~{\rm GeV}$) are required to match experimental results. This is because under these conditions, the effective annihilation cross-section of DM can be significantly lower than that predicted by the MSSM for the same mass, thereby increasing the relic density.

\item As $|m_{\tilde{\chi}_1^0}|$ gradually increases, the $2 \sigma$ upper limits of $\lambda$, $d^\prime$, and $N_{15}$ all rise significantly, from $0.12$, $0.25$, and $-0.7$ to $0.32$, $0.7$, and $-0.28$, respectively. This occurs because as $|m_{\tilde{\chi}_1^0}|$ increases, DM can achieve the experimentally measured relic abundance through the co-annihilation of Higgsino-dominated particles alone, making the mixing between Higgsino and Singlino no longer a necessary condition for predicting the correct relic density. Consequently, the relic density’s dependence on $N_{15}$ weakens, allowing it to vary over a wider range (for instance, $−0.6 \lesssim N_{15} \lesssim −0.28$ when the DM mass is $1000~{\rm GeV}$). According to Eq.~(\ref{mixing angle}), both $\lambda$ and $d^\prime$ can take larger values in this case. In comparison, as $|m_{\tilde{\chi}_1^0}|$ increases from $700~{\rm GeV}$, the $2 \sigma$ upper limit of $\Delta m$ shows only a slight increase, stabilizing around $25~{\rm GeV}$. This behavior arises because $\Delta m$ appears only in the exponential term of $\sigma_{eff}$ in Eq.~(\ref
{eff-cross-section}). If we perform a Taylor expansion of Eq.~(\ref{eff-cross-section}), the leading-order contribution of the co-annihilations to $\sigma_{\rm eff}$, usually acting as the dominant component of the total cross-section,  is approximately independent of DM mass. This fact suggests that the relic density and its related $\Delta m$ are also largely independent of DM mass.
    
\begin{figure*}[t]
  \centering
  \resizebox{1.0 \textwidth}{!}{
    \includegraphics{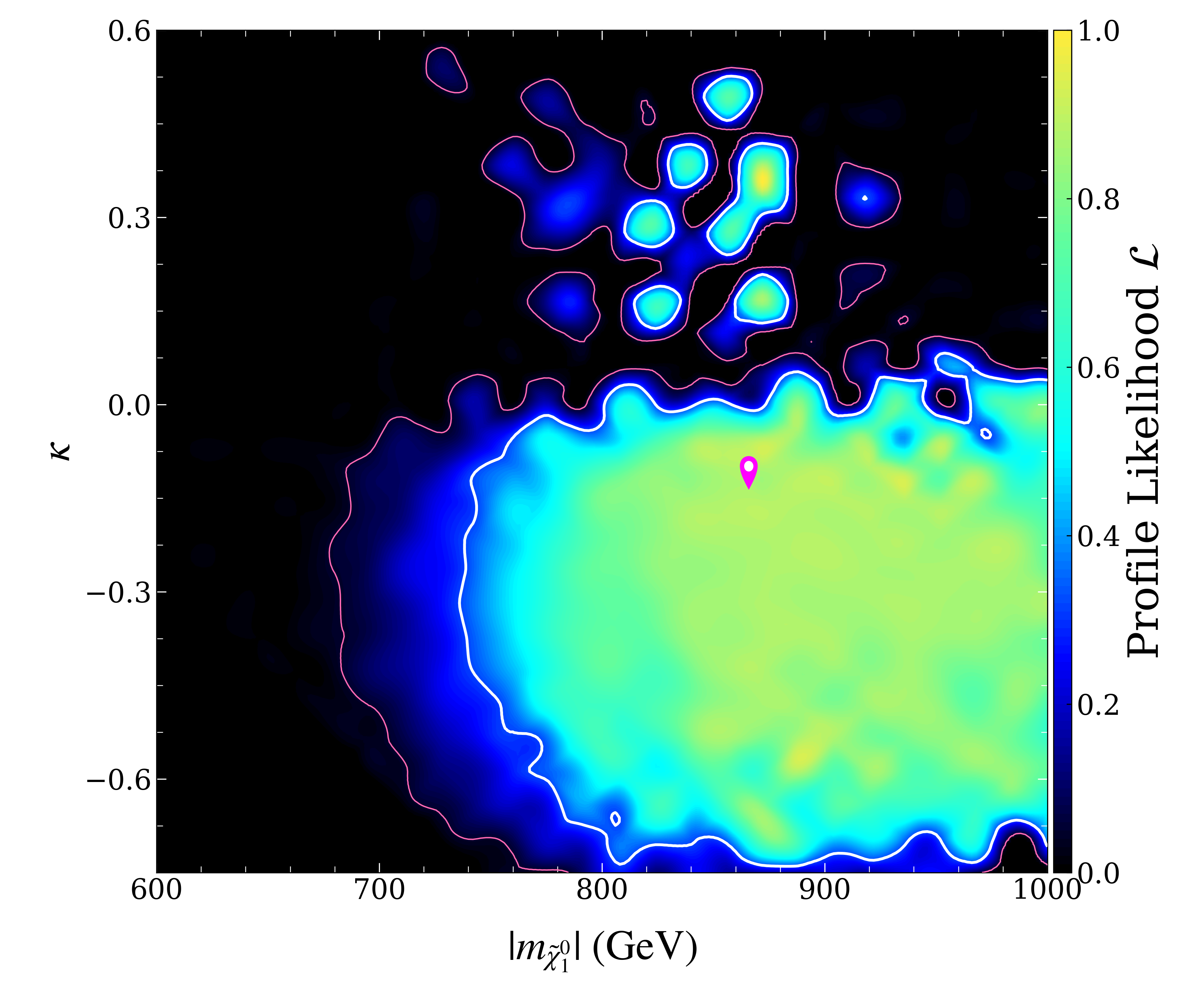}
    \includegraphics{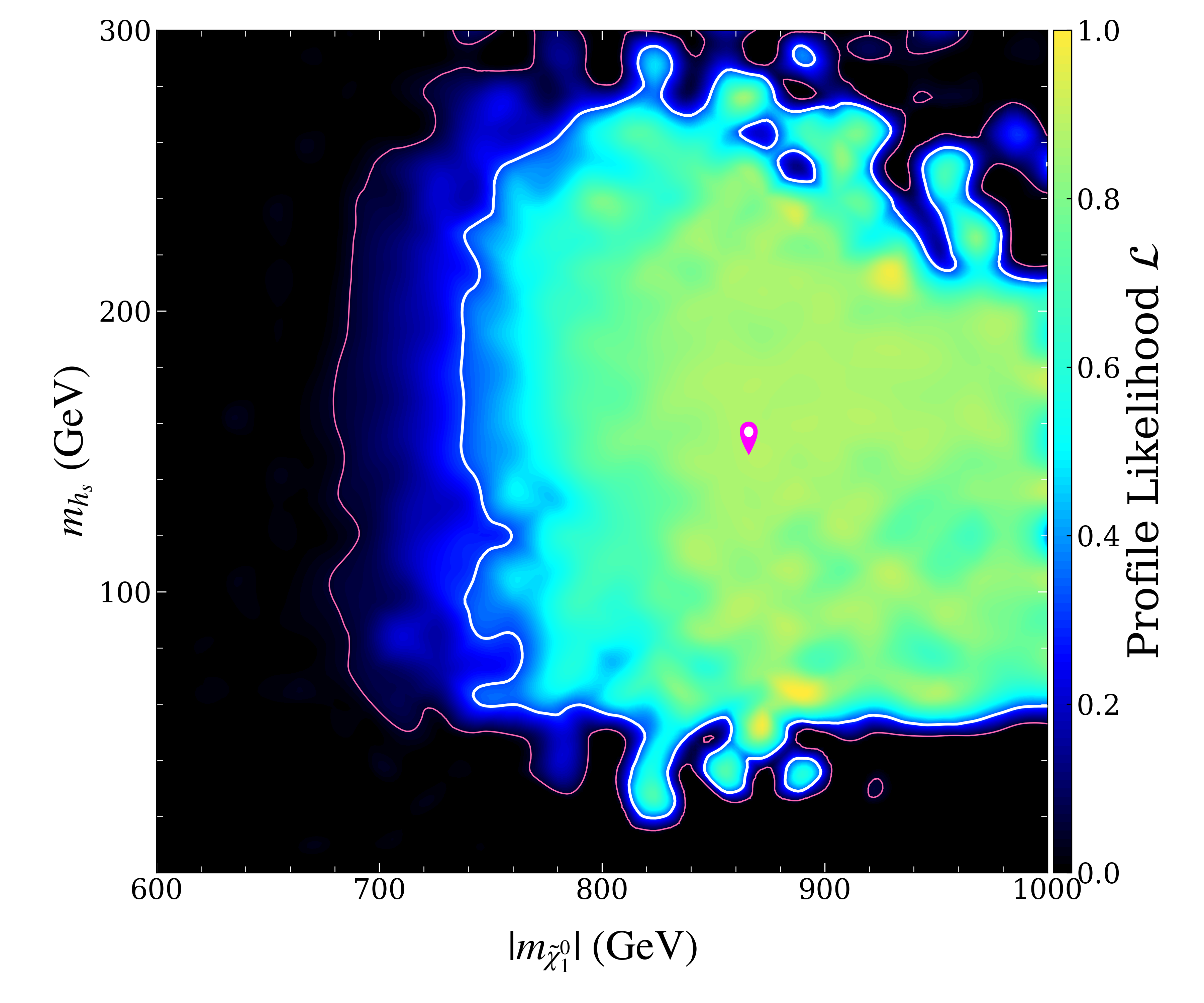}  
  }
  
  \resizebox{1.0 \textwidth}{!}{
    \includegraphics{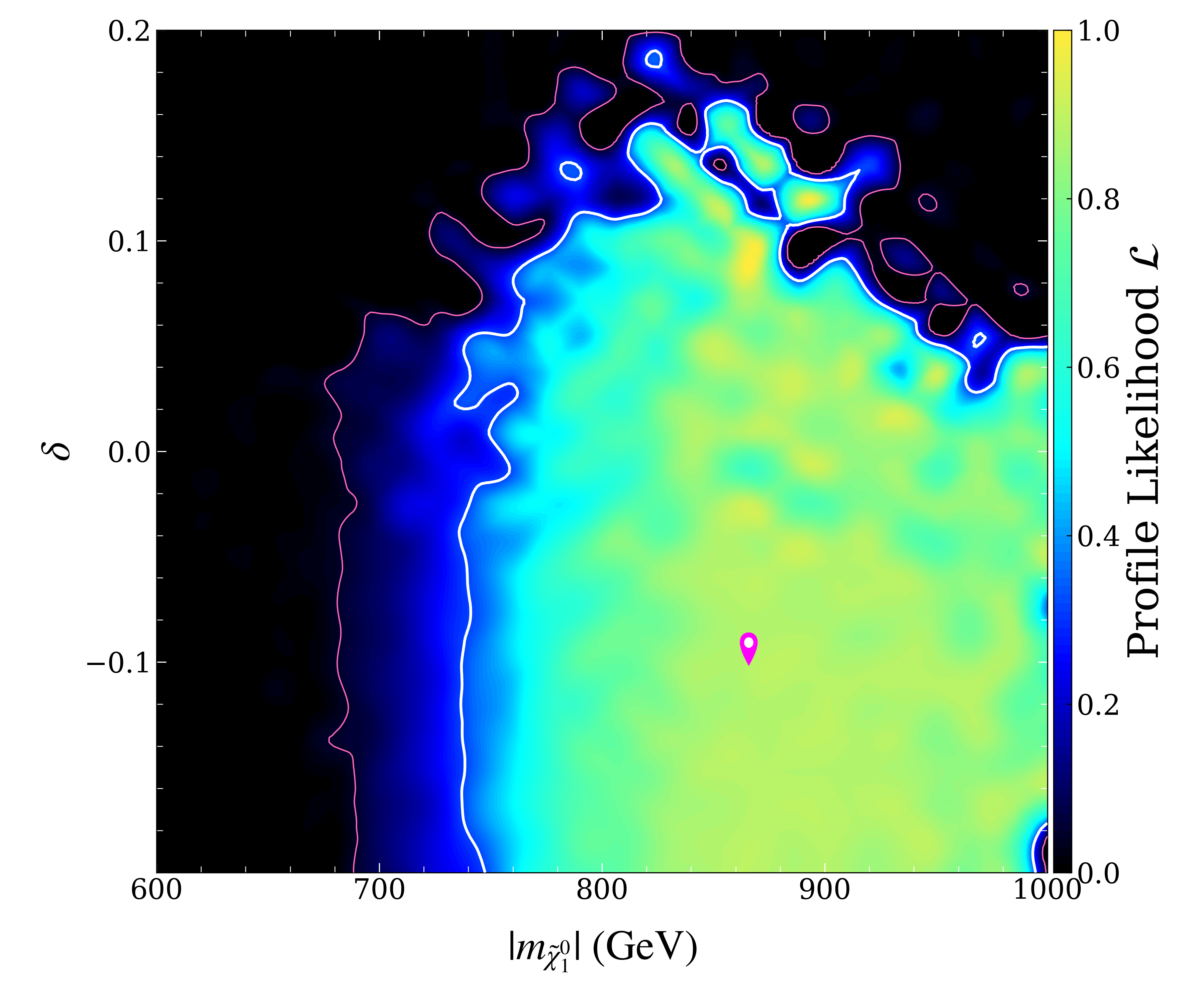}
    \includegraphics{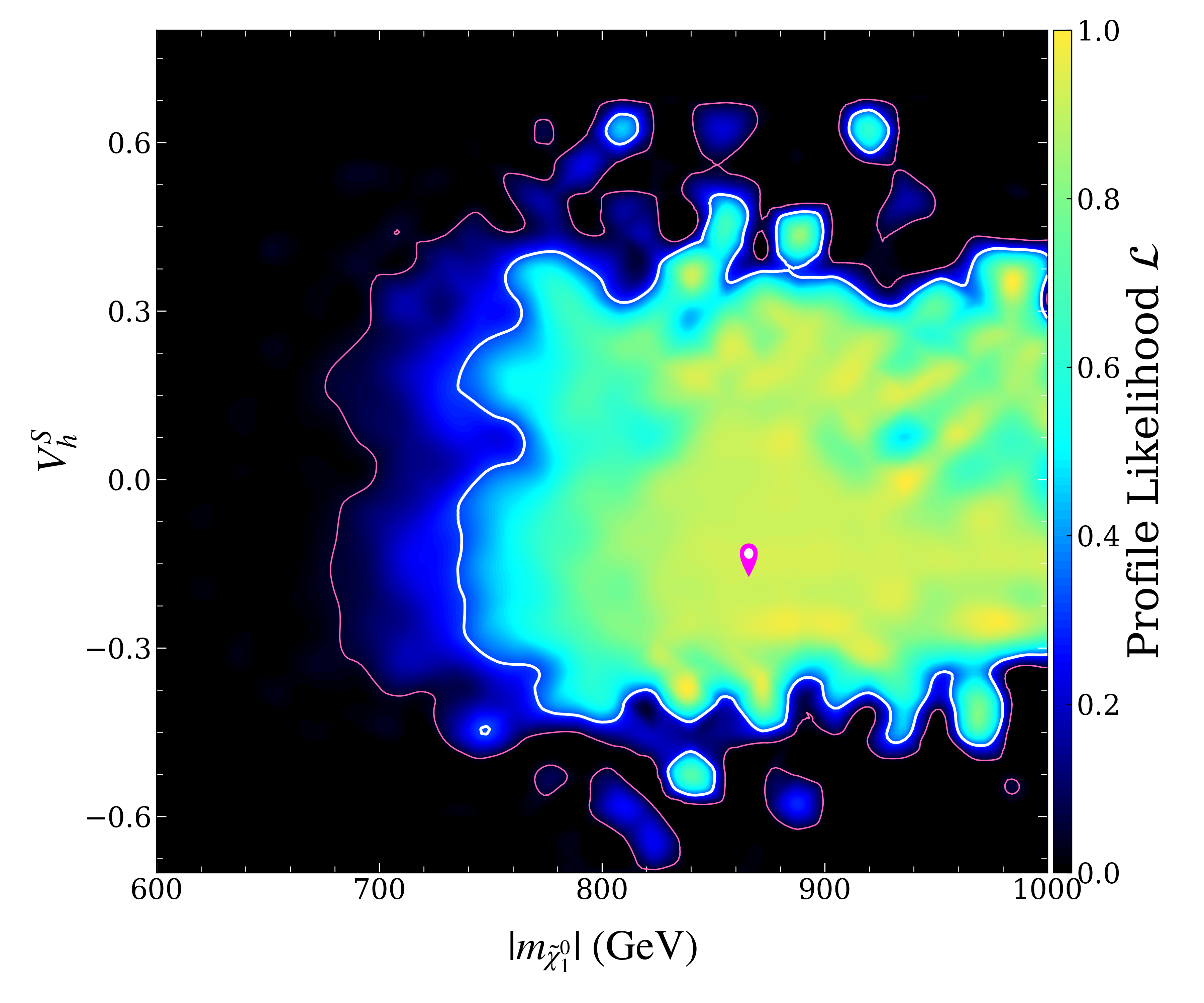}
  }
  \vspace{-0.5cm}
  \caption{Same as Fig~\ref{Fig1}, but for the profile likelihood projected onto  the $|m_{\tilde{\chi}_1^0}|-\kappa$,  $|m_{\tilde{\chi}_1^0}|-m_{h_s}$,  $|m_{\tilde{\chi}_1^0}|-\delta$, and  $|m_{\tilde{\chi}_1^0}|-V_h^S$ planes, respectively. \label{Fig2}}
\end{figure*}

\item The yellow regions in Fig.~\ref{Fig1} correspond to relatively small $\chi^2$ values, indicating that these areas can well explain experimental results. The PL map on the $|m_{\tilde{\chi}_1^0}|-\lambda$ plane shows that $\lambda$ in this region is concentrated within a narrow range. This is because $\lambda$ not only affects DM relic density through $N_{15}$ and $\Delta m$, but also influences Higgs physics (see the Higgs mass matrix in Eq~(\ref{CP-even-mass-matrix})) and DM-nucleon scattering in more complex ways. As a result, $\lambda$ is a critical parameter strongly constrained by various experiments, with current LZ experimental results favoring tiny $\lambda$ values. Fig.~\ref{Fig1} also shows that $d^\prime$ and $N_{15}$ can span a wider range, reflecting the fact that, for larger DM masses, neither $d^\prime$ nor $N_{15}$ is crucial to DM physics. Moreover, $\Delta m$ is similarly confined to a narrow range in the figure. This is because $\Delta m$ affects the relic density exponentially, and large variations in $\Delta m$ would lead to rapid changes in the relic abundance, which contradicts experimental results.
\end{itemize}

Next, we discuss how direct detection experiments constrain the parameter space. From Eq.~(\ref{SIA}), we know that the SI scattering cross-section between DM and nucleons depends on parameters $\lambda$, $N_{15}$, $\kappa$, $m_{h_s}$, and $V_h^{S}$. Fig.~\ref{Fig2} depicts the PL distributions for the $m_{\tilde{\chi}_1^0}-\kappa$, $m_{\tilde{\chi}_1^0}-m_{h_s}$, $m_{\tilde{\chi}_1^0}-\delta$,  and $m_{\tilde{\chi}_1^0}-V_h^S$ planes, where parameter $\delta$ influences $V_h^S$ through Eq.~(\ref{Approximations}). The results show that $\kappa$, $m_{h_s}$, $\delta$, and $V_h^S$ can vary across relatively large regions, primarily due to two reasons:
\begin{itemize}
\item Eq.~(\ref{SIA}) indicates that the leading-order contribution to the SI scattering cross section is proportional to $\lambda^4$. When $\lambda$ is sufficiently small, the cross-section can be drastically reduced, ensuring compatibility with the LZ experiment and allowing other parameters to be larger.
\item Depending on the signs of $N_{15}$, $\kappa$, and $V_h^S$, different terms in Eq.~(\ref{SIA}) may cancel each other, further reducing the scattering cross-section between DM and nucleons. When $\kappa$ takes large values and $h_s$ is light, the influence of the singlet-dominated Higgs particle on the scattering cross-section increases, which can further enhances this cancellation effect. Combining the results from Fig.~\ref{Fig1} and Fig.~\ref{Fig2}, one sees that negative values of $N_{15}$, $\kappa$, $\delta$, and $V_h^S$ tend to strengthen these cancellations, thus aligning more easily with LZ experimental results.
\end{itemize} 

Furthermore, we found that since the SD scattering cross-section between DM and nucleons is suppressed, its constraints on the parameter space are generally weaker compared to SI scattering when considering the LZ data in 2022. However, in certain special cases, it may also provide more stringent constraints.

\begin{figure*}[t]
  \centering

  \resizebox{1.0 \textwidth}{!}{
    \includegraphics{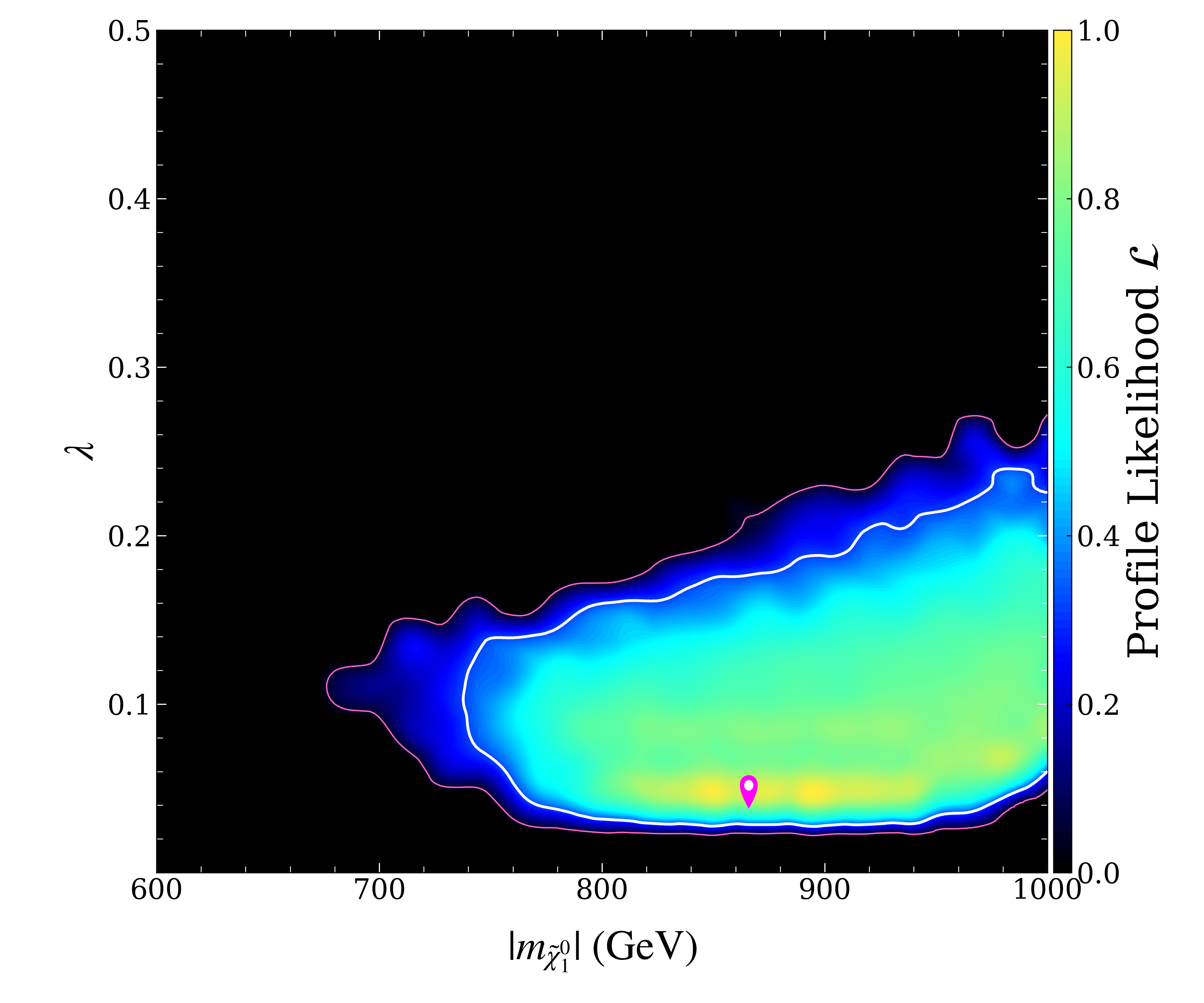}
    \includegraphics{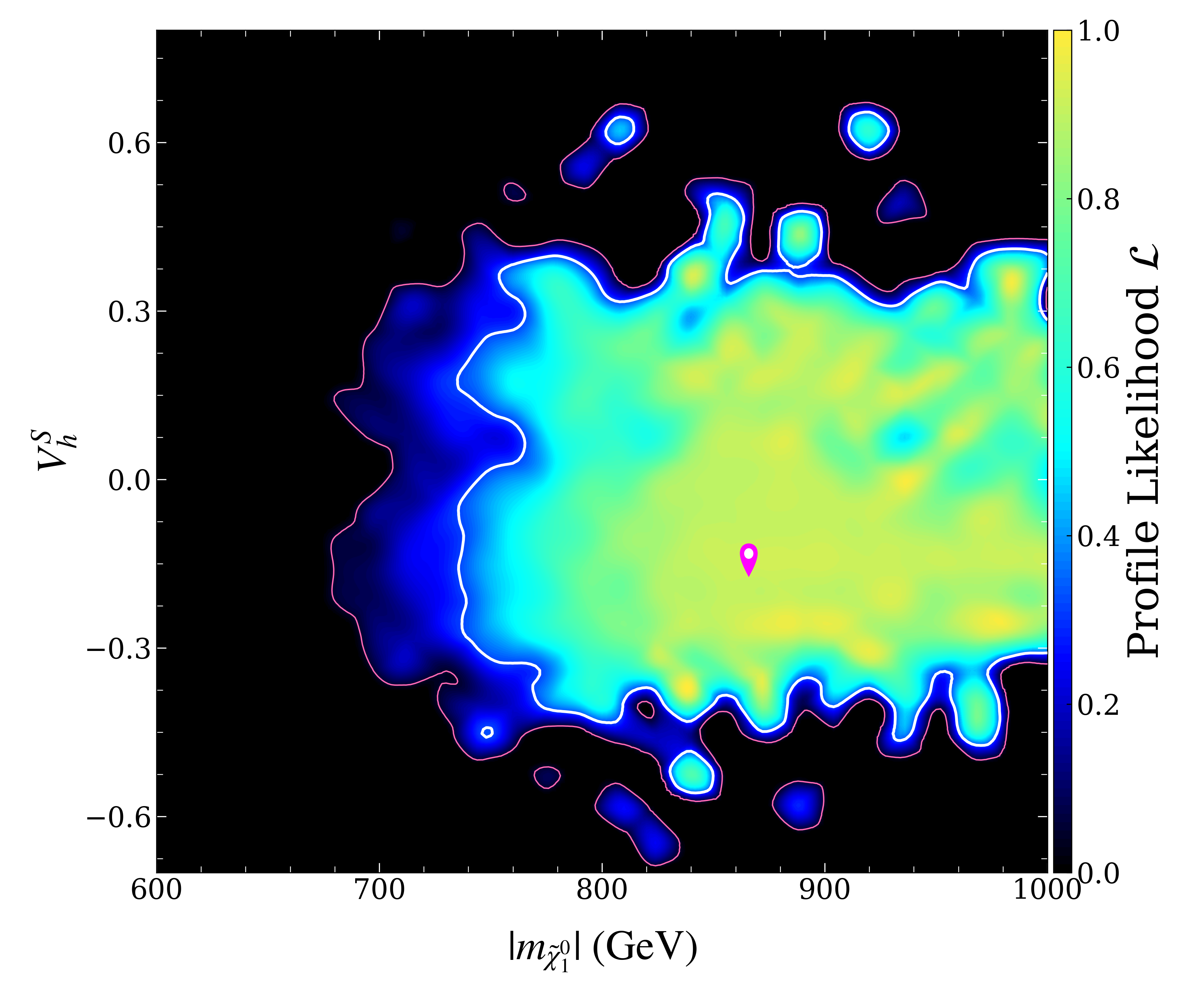}
  }
  \vspace{-0.5cm}

  \caption{Profile likelihood maps of $\cal{L}$, projected onto the $|m_{\tilde{\chi}_1^0}|-\lambda$ planes (upper panels) and the $|m_{\tilde{\chi}_1^0}|-V_h^S$ planes. In obtaining this figure, we updated the samples in Fig.~\ref{Fig1} by including the constraints from the 2024 LZ results.
  \label{Fig3}}
\end{figure*}

\begin{figure*}[t]
  \centering
  \resizebox{1.0 \textwidth}{!}{
   \includegraphics{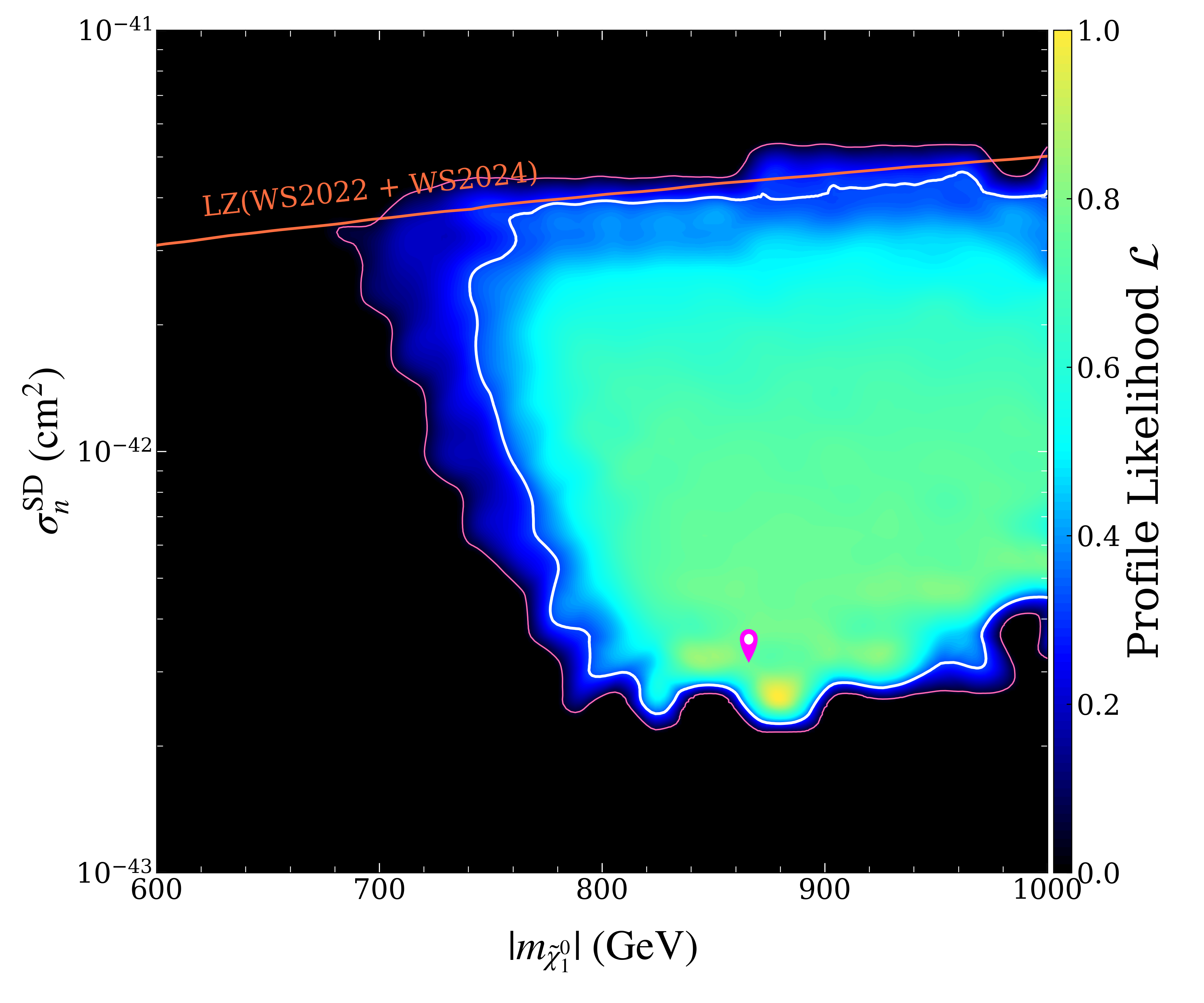}
    \includegraphics{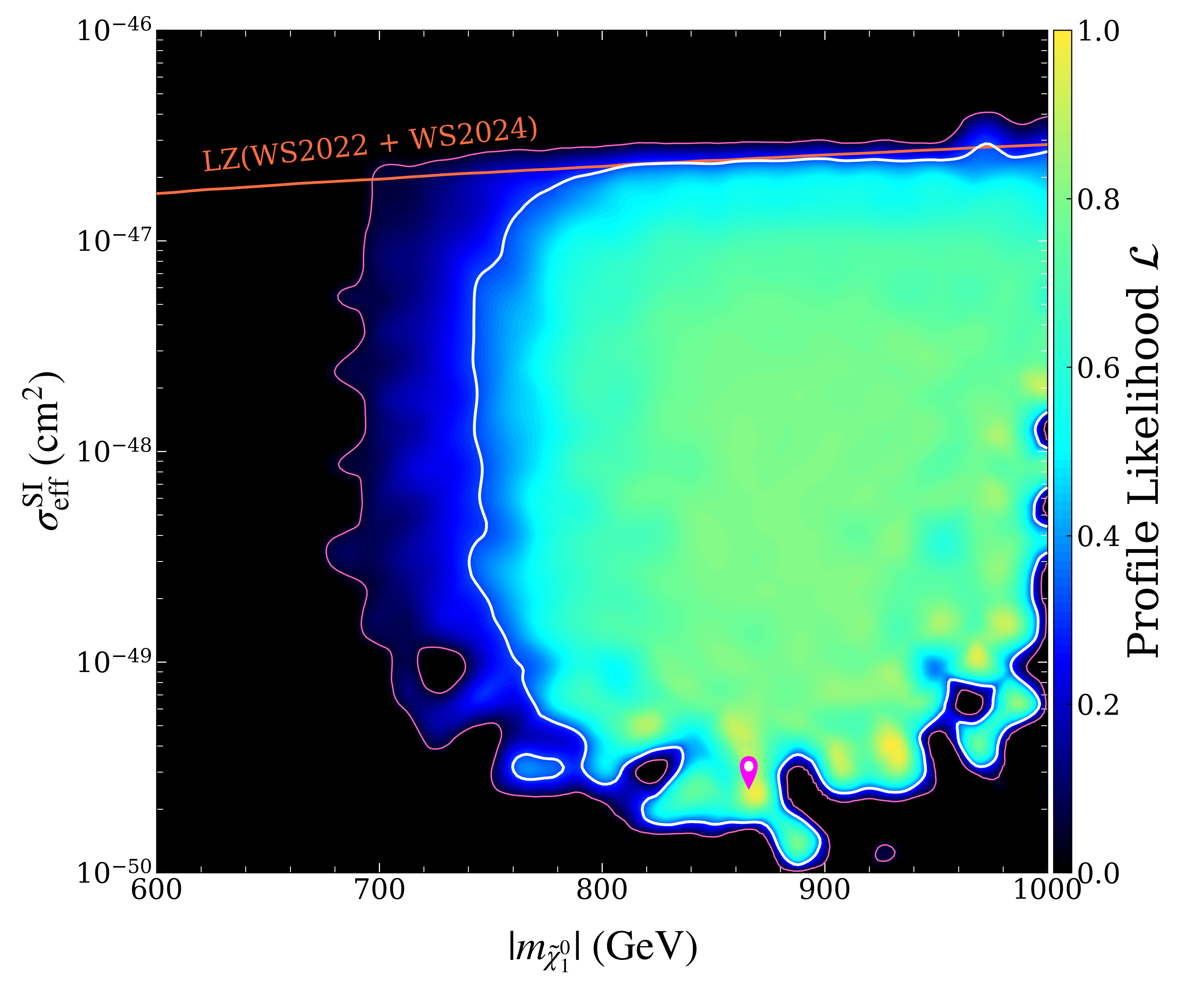}
  }
  \vspace{-0.5cm}
  \caption{Profile likelihoods of $\sigma^{\rm SD}_n$ and $\sigma^{\rm SI}_{\rm eff}$ as functions the DM mass. All displayed points satisfy the complete set of updated experimental constraints, reflecting current status of DM interactions with nucleon in the GNMSSM. \label{Fig4}}
\end{figure*}

\begin{table}[htbp]  
  \centering  
  \resizebox{\textwidth}{!}{  
  \begin{tabular}{lr@{}ll|lr@{}ll}   
      \toprule  
      \multicolumn{4}{c|}{\textbf{Point P1}} & \multicolumn{4}{c}{\textbf{Point P2}} \\ \midrule  
      Parameter     & &  & Value  & Parameter & &  & Value  \\ \midrule  
      $\tan{\beta}$ & & & $21.96$ & $\tan{\beta}$ & & & $29.86$ \\
      $\lambda$, $\kappa$     & & & $0.037,-0.13$ & $\lambda$, $\kappa$ & & & $0.009,-0.029$ \\
      $d^{\prime}$, $\delta$ & & & $0.21,-0.10$  & $d^{\prime}$, $\delta$ & & & $1.80,0.17$ \\
      $v_s$,  $\mu_{\rm tot}$ & & & $600.0,853.2$ & $v_s$,  $\mu_{\rm tot}$ & & & $600.0,1115.2$  \\
      $m_A$, $m_B$, $m_C$ & & & $3000,194.3,400$ & $m_A$,$m_B$,$m_C$ & & & $3000,135.4,400$ \\
      $M_1$,$M_2$ & & & $2000,2000$ & $M_1$,$M_2$ & & & $6000,6000$ \\ \midrule

      Particle      & & & Mass spectrum & Particle  & & & Mass spectrum \\         
      $\tilde{\chi}_1^0$,$\tilde{\chi}_2^0$, $\tilde{\chi}_3^0$ & & & $-865.9,866.1,-875.5$ & $\tilde{\chi}_1^0$,$\tilde{\chi}_2^0$, $\tilde{\chi}_3^0$ & & & $1136.9,-1137.3,-1316.1$ \\
      $\tilde{\chi}_4^0$,  $\tilde{\chi}_5^0$ & & & $2007.4, 2023.3$ & $\tilde{\chi}_4^0$, $\tilde{\chi}_5^0$ & & & $5940.1,6064.5$  \\
      $\tilde{\chi}_1^{\pm}$,  $\tilde{\chi}_2^{\pm}$ & & & $868.2,2023.2$ & $\tilde{\chi}_1^{\pm}$, $\tilde{\chi}_2^{\pm}$ & & & $1137.5,5940.2$  \\
      $h_s$, $h$, $H$  & & & $148.6,124.6,3242.2$ & $h_s$, $h$, $H$  & & & $116.2,124.5,4242.8$ \\ \midrule   
      
      Rotation matrix     & & & Element value & Rotation matrix & & & Element value \\      
      \multicolumn{2}{l}{$V_{h_s}^S$,$V_{h_s}^{SM}$,$V_{h}^S$,$V_{h}^{SM}$} & \multicolumn{2}{l|}{$-98,\ 17,\ -17,\ -98$} &   
      \multicolumn{2}{l}{$V_{h_s}^S$,$V_{h_s}^{SM}$,$V_{h}^S$,$V_{h}^{SM}$} & \multicolumn{2}{l}{$96,\ -25,\ 24,\ 96$} \\
      
      \multicolumn{2}{l}{$N_{11}$,$N_{12}$,$N_{13}$,$N_{14}$,$N_{15}$} & \multicolumn{2}{l|}{$-1,\ 1,\ 50,\ 51,\ -70$} &   
      \multicolumn{2}{l}{$N_{11}$,$N_{12}$,$N_{13}$,$N_{14}$,$N_{15}$} & \multicolumn{2}{l}{$1,\ -1,\ 71,\ -70,\ 0$} \\
      
      \multicolumn{2}{l}{$N_{21}$,$N_{22}$,$N_{23}$,$N_{24}$,$N_{25}$} & \multicolumn{2}{l|}{-3,\ 5,\ -71,\ 71,\ 0} &   
      \multicolumn{2}{l}{$N_{21}$,$N_{22}$,$N_{23}$,$N_{24}$,$N_{25}$} & \multicolumn{2}{l}{0,\ -1,\ -71,\ -71,\ -1} \\
      
      \multicolumn{2}{l}{$N_{31}$,$N_{32}$,$N_{33}$,$N_{34}$,$N_{35}$} & \multicolumn{2}{l|}{1,\ -1,\ -50,\ -50,\ -71} &   
      \multicolumn{2}{l}{$N_{31}$,$N_{32}$,$N_{33}$,$N_{34}$,$N_{35}$} & \multicolumn{2}{l}{0,\ 0,\ 1,\ -4,\ 99} \\
      
      \multicolumn{2}{l}{$N_{41}$,$N_{42}$,$N_{43}$,$N_{44}$,$N_{45}$} & \multicolumn{2}{l|}{-47,\ -88,\ -1,\ 3,\ 0} &   
      \multicolumn{2}{l}{$N_{41}$,$N_{42}$,$N_{43}$,$N_{44}$,$N_{45}$} & \multicolumn{2}{l}{0,\ 99,\ 0,\ -1,\ 0} \\
      
      \multicolumn{2}{l}{$N_{51}$,$N_{52}$,$N_{53}$,$N_{54}$,$N_{55}$} & \multicolumn{2}{l|}{88,\ -47,\ -2,\ 5,\ 0} &   
      \multicolumn{2}{l}{$N_{51}$,$N_{52}$,$N_{53}$,$N_{54}$,$N_{55}$} & \multicolumn{2}{l}{-99,\ 0,\ 0,\ -1,\ 0} \\ \midrule  
      
      Primary annihilation & & & Fraction [\%]               & Primary annihilation & & & Fraction [\%] \\
     $\tilde{\chi}_2^0  \tilde{\chi}_1^{\pm}\to d \bar{u} / \bar{d} u$ & & &~~$8.40$ & $\tilde{\chi}_1^0 \tilde{\chi}_2^{0} \to t \bar{t} $ & & &~~$6.90$ \\
     $\tilde{\chi}_2^0 \tilde{\chi}_1^{\pm} \to s \bar{c} / \bar{s} c $ & & &~~$8.40$ & $\tilde{\chi}_1^0 \tilde{\chi}_1^{\pm} \to u \bar{d} / \bar{d} u$ & & &~~$6.73$ \\
     $\tilde{\chi}_1^0  \tilde{\chi}_1^{\pm} \to   d \bar{u} / \bar{u} d $ & & &~~$4.36$ & $\tilde{\chi}_1^0 \tilde{\chi}_1^{\pm} \to s \bar{c} / \bar{c} s $ & & &~~$6.73$ \\
     $\tilde{\chi}_1^0 \tilde{\chi}_1^{\pm} \to  c \bar{s} / \bar{c} s$ & & &~~$4.36$ & $\tilde{\chi}_2^0 \tilde{\chi}_1^{\pm} \to d \bar{u} / \bar{d} u $ & & &~~$6.43$ \\
     $\tilde{\chi}_2^0 \tilde{\chi}_1^{\pm} \to  e \bar{\nu_e} / \bar{\nu_e} e$ & & &~~$2.80$ & $\tilde{\chi}_2^0 \tilde{\chi}_1^{\pm} \to s \bar{c} / \bar{s} c $ & & &~~$6.43$ \\
     $\tilde{\chi}_1^0 \tilde{\chi}_1^{\pm} \to  \mu \bar{\nu_{\mu}} / \bar{\nu_{\mu}} \mu$ & & &~~$2.80$ & $\tilde{\chi}_2^{+} \tilde{\chi}_1^{-} \to t \bar{t} $ & & &~~$5.41$ \\
     $\tilde{\chi}_1^{+} \tilde{\chi}_1^{-} \to  W^{+} W^{-}$ & & &~~$2.67$ & $\tilde{\chi}_2^{0} \tilde{\chi}_1^{\pm} \to b \bar{t} / t \bar{b} $ & & &~~$4.34$ \\
     $\tilde{\chi}_2^0 \tilde{\chi}_1^{\pm} \to  \tau \bar{\nu_{\tau}} / \bar{\nu_{\tau}} \tau$ & & &~~$2.59$ & $\tilde{\chi}_1^{+} \tilde{\chi}_1^{-} \to b \bar{t} / t \bar{b}  $ & & &~~$4.22$ \\ \midrule
     
      DM observable     & & & Value & DM observable & & & Value \\   
      $\Omega h^2$ & & & $0.120$ & $\Omega h^2$ & & & $0.118$ \\
      $\sigma^{\rm SI}_{p,n}/(10^{-50} \text{cm}^2)$ & & & $9.67,0.25 $ & $\sigma^{\rm SI}_{p,n}/(10^{-47} \text{cm}^2)$ & & & $2.25,2.31$ \\
      $\sigma^{\rm SD}_{p,n}/(10^{-43} \text{cm}^2)$ & & & $4.12,3.14 $ & $\sigma^{\rm SD}_{p,n}/(10^{-45} \text{cm}^2)$ & & & $2.67,2.05$ \\    \bottomrule  
  \end{tabular}  
  }  
  \caption{Detailed information for two benchmark points compatible with all experimental measurements. Point P1 represents the best-fit point from Fig.~\ref{Fig3}, while P2 corresponds to the MSSM limit of the GNMSSM, characterized by a particularly small $\lambda$ value. All mass-dimensional parameters are expressed in GeV, and elements of the rotation matrices $V$ and $N$ are presented in units of $10^{-2}$. \label{Tab2}}  
\end{table}  

During the revision of this study, we noted that the 2024 LZ results had just been published~\cite{LZ:2024zvo}, which significantly improved the sensitivity of DM-nucleon scattering cross-section by nearly one order of magnitude compared to the results from 2022. This enhancement necessitated updating our analysis to incorporate the new constraints. After applying these refinements, our sample size decreased from 276,077 points obtained in Fig.~{\ref{Fig1} and ~\ref{Fig2} to 30,299 points. Again, this substantial reduction demonstrates the remarkable constraining power of recent experimental advances in limiting the viable parameter points of the Higgsino DM scenario. However, despite this significant reduction, most of the two-dimensional PL maps in Fig.~\ref{Fig1} and Fig.~\ref{Fig2} remained largely unchanged, with notable exceptions occurring only in the  $|m_{\tilde{\chi}_1^0}|-\lambda$  and $|m_{\tilde{\chi}_1^0}|-V_h^S$ planes. This selective sensitivity highlights a key insight into the model: among all parameters, $\lambda$ (and consequently $V_h^S$) alone plays particularly crucial roles in Higgs and DM phenomenology.

To clearly illustrate the impacts of the 2024 LZ results, we present in Fig.~\ref{Fig3} the PL maps of $\cal{L}$ projected onto the $|m_{\tilde{\chi}_1^0}|-\lambda$ and $|m_{\tilde{\chi}_1^0}|-V_h^S$ planes. The left panels demonstrate that the $2\sigma$ upper bound on $\lambda$ decreases from approximately $0.32$ in Fig.~\ref{Fig1} to $0.28$. Remarkably, the most favored region for $\lambda$, highlighted by the yellow band, remains stable around $0.04$, consistent with our previous findings. Turning to $V_h^S$, the allowed region contracts significantly, particularly near the boundaries, as clearly demonstrated by comparing the bottom-right panel of Fig.~\ref{Fig2} with the right panels of Fig.~\ref{Fig3}. Nevertheless, even under these increasingly stringent constraints, our analysis reveals that DM can still be as light as approximately $670~{\rm GeV}$ while remaining fully compatible with all experimental measurements.

We also present in Fig.\ref{Fig4} the PL map of $\cal{L}$ on the $|m_{\tilde{\chi}_1^0}|-\sigma^{\rm SD}_n$ and $|m_{\tilde{\chi}_1^0}|-\sigma^{\rm SI}_{\rm eff}$ planes, considering all samples that satisfy the updated constraints. For reference, the latest LZ exclusion bounds from Ref.~\cite{LZ:2024zvo} are also shown. It can be seen that  $\sigma^{\rm SD}_n$ can be reduced to about $2 \times 10^{-43}~{\rm cm^2}$ for $\lambda \simeq 0.02$, roughly one order of magnitude below the current LZ limits. Even more dramatically, $\sigma^{\rm SI}_{\rm eff}$ can reach values as low as $10^{-50}~{\rm cm^2}$, which lies three orders of magnitude below the corresponding LZ limits and two orders of magnitude below the neutrino floor. The reason behind these disparities is that as indicated by Eq.~(\ref{SD-approximation}),  $\sigma^{\rm SD}_n$ is suppressed only by a factor of $\lambda^4 v^4/(d^2 \mu_{\rm tot}^4)$, whereas $\sigma^{\rm SI}_{\rm eff}$ depends on multiple parameters of the GNMSSM, each contribution being suppressed by at least $\lambda^4$ with the added possibility of cancellations among different terms. Our analysis further indicates important implications for future experiments: if upcoming DM direct detection experiments improve current sensitivities by just one order of magnitude without finding any evidence of DM, the parameter $\lambda$ would be constrained to values significantly smaller than $0.02$. This constraint would primarily arise from limitations on $\sigma^{\rm SD}_n$ rather than on $\sigma^{\rm SI}_{\rm eff}$. Consequently, the Bayesian evidence for the Higgsino DM scenario within the GNMSSM framework would be substantially reduced.

\subsection{Benchmark Points}

Before completing the presentation of the results, we show in Table~\ref{Tab2} detailed information for two benchmark points that satisfy all experimental constraints. Point P1 corresponds to the best-fit point obtained by minimizing the $\chi^2$ function specified in footnote 1. It features substantial Higgsino-Singlino mixing, enabling a sub-TeV $\tilde{\chi}_1^0$ to achieve the measured DM relic abundance primarily through co-annihilation processes involving Higgsino-dominated states.
By contrast, point P2 represents the MSSM limit within the GNMSSM framework, illustrating a viable pure Higgsino DM scenarios. It is characterized by an extremely small $\lambda$, which explicitly suppresses the Higgsino-Singlino mixing. This fundamental difference in mixing significantly alters the relative importance of various annihilation channels in the two scenarios. For instance, when examining the effective annihilation cross-section $\sigma_{eff}$ formulated in Eq.~(\ref{eff-cross-section}), we find that the $\tilde{\chi}_1^0 \tilde{\chi}_2^0 \to t \bar{t}$ process contributes only $1.1\%$ to $\sigma_{eff}$ for Point P1, whereas this contribution increases to $6.9\%$ for Point P2. Consequently, a heavier $\tilde{\chi}_1^0$ mass of approximately $1.1~{\rm TeV}$ is required for Point P2 to attain the correct relic abundance via Higgsino co-annihilation mechanisms.

Aside from these evident differences, another crucial distinction between the two points lies in the gaugino masses: $M_1$ and $M_2$ in P1 are significantly smaller than those in P2, yet both configurations remain compatible with LZ results. This compatibility arises because the effects of Higgsino-Singlino mixing on DM-nucleon scattering cross-sections can effectively cancel against those from Gaugino-Higgsino mixing, allowing for relatively light gauginos around 2 TeV to remain consistent with the direct detection constraints.  To illustrate the importance of this cancellation mechanism, consider that if we set  $M_1 = M_2 = 2~{\rm TeV}$ while keeping other parameters of P2 unchanged, the predicted cross-sections would increase drastically: $\sigma^{\rm SI}_p = 6.8 \times 10^{-46}~{\rm cm^2}$,  $\sigma^{\rm SI}_n = 7.0 \times 10^{-46}~{\rm cm^2}$,  $\sigma^{\rm SD}_p = 4.7 \times 10^{-43}~{\rm cm^2}$, and $\sigma^{\rm SD}_n = 3.6 \times 10^{-43}~{\rm cm^2}$. All of these values exceed current LZ upper bounds by more than one order of magnitude. These findings underscore the central conclusion of this work: the GNMSSM provides a significantly broader and more flexible parameter space than the MSSM for accommodating viable Higgsino-dominated DM scenarios.

\section{Other Issues\label{Other_Issues}}

This section addresses other important aspects of our study, including its relationship to previous work and the prospects for future experimental detection.

\subsection{Study Context and Novel Contributions}

As established in the introduction, the properties of Higgsino dark matter have been extensively investigated within the MSSM framework, where it is well-established that thermal freeze-out reproduces the observed relic abundance only when the DM mass is approximately  $1.1~{\rm TeV}$. Most recently, Ref.~\cite{Martin:2024ytt} applied the LZ-2024 results to constrain this scenario, finding that $|M_1| \gtrsim 1.3~{\rm TeV}$ and $|M_2| \gtrsim 2.1~{\rm TeV}$ are required, with opposite signs for $M_1$ and $M_2$  being preferred.

In contrast, investigations of DM in the NMSSM have predominantly focused on Bino- or Singlino-dominated scenarios, while systematic analyses of Higgsino-dominated DM remain relatively scarce. To our knowledge, only a limited number of works have addressed this specific direction. We summarize the relevant previous studies below:

\begin{itemize}
\item Ref.~\cite{Cao:2016nix} examined Higgsino DM within the $Z_3$-NMSSM, demonstrating that for $120~{\rm GeV} \leq M_1 \leq 500~{\rm GeV}$ and  $165~{\rm GeV} \leq M_2 \leq 1000~{\rm GeV}$, the DM with a mass ranging from from $65~{\rm GeV}$ to $85~{\rm  GeV}$ could simultaneously satisfy relic abundance requirements and the direct detection constraints available at that time, provided sizable Higgsino-Singlino mixing was present. However, this parameter region has since been excluded by the latest LZ and LHC experimental results.
    
\item Ref.~\cite{Mummidi:2018myd} investigated the Split NMSSM scenario, where supersymmetry breaking occurs above  $10^{14}~{\rm GeV}$, leaving only one Higgsino pair and a Singlino at an intermediate scale in the low-energy spectrum. Using effective field theory estimates, the authors demonstrated that when the Singlino mass lies below approximately $10^{8}~{\rm GeV}$, Higgsino–Singlino mixing induces a mass splitting between neutral Higgsinos exceeding $200~{\rm MeV}$, effectively rendering them a pseudo-Dirac fermion that can evade direct detection limits. The paper's primary focus was on vacuum stability and Higgs physics. Its emphasis is therefore different from ours, although our analysis addresses some of its key considerations from a different perspective.
      
\item Ref.~\cite{Wang:2024ozr} studied Higgsino-dominated DM in the semi-constrained $Z_3$-NMSSM through parameter space scans, but imposed only an upper bound on the relic abundance rather than requiring exact matching to observations. As evident from its Fig. 1, most parameter points do not exhibit significant Higgsino–Singlino mixing effects, making their results less relevant to the scenario we investigate here.
\end{itemize}    

The present work makes several distinct contributions that advance our understanding beyond these previous studies. First, we systematically highlight the crucial role of Higgsino-Singlino mixing in determining the properties of Higgsino-dominated DM, demonstrating how this mixing can significantly modify both the mass spectrum and interaction strengths. Second, we extend and update the findings of Ref.~\cite{Cao:2016nix} within the more general GNMSSM framework, which provides additional flexibility compared to the $Z_3$-invariant model. By incorporating the latest LZ-2024 data, we identify viable parameter regions that satisfy all current experimental constraints—regions that differ substantially from those reported in earlier studies. 
Most importantly, we derive comprehensive analytic expressions for the effects of Higgsino-Singlino mixing on both the mass spectrum and DM-nucleon scattering cross sections. To the best of our knowledge, such detailed analytic formulations have not appeared in the literature previously. These expressions serve to make our numerical results more transparent, physically interpretable, and readily reproducible, providing a valuable theoretical framework for future investigations.

\subsection{Future Detection Prospects}

Our results reveal that the Higgsino-dominated DM mass can be as low as $670~{\rm GeV}$, with mass splittings between $\tilde{\chi}_1^0$ and $\tilde{\chi}_1^{\pm}$ and $\tilde{\chi}_2^0$  varying from a few GeV to around $20~{\rm GeV}$, corresponding to compressed spectra scenarios in SUSY search at the LHC. While these scenarios present challenges for current experiments due to small mass splittings and soft decay products, several future experimental approaches offer promising detection prospects:

\begin{itemize}
    \item \textbf{Collider Search}  

   The High-Luminosity LHC (HL-LHC) operating at  $\sqrt{s} = 14~{\rm TeV}$ with an integrated luminosity of $3000~{\rm fb}^{-1}$ can probe compressed spectra scenarios up to $600~{\rm GeV}$ through the disappearing track signature arising from the chargino's characteristic lifetime of $0.1~{\rm ns}$ ($c \tau \approx 3~{\rm cm}$)~\cite{Mlynarikova:2023bvx,Fukuda:2017jmk,ATL-PHYS-PUB-2022-018}. This reach extends to an exclusion limit of $850~{\rm GeV}$ or a discovery reach of $600~{\rm GeV}$ for a future $100~{\rm TeV}$ collider~\cite{Low:2014cba}. Furthermore, a future $e^+ e^-$ collider operating at $\sqrt{s} = 3~{\rm TeV}$  would provide the most precise measurements in a clean environment, allowing for the discovery of Higgsinos up to $1~{\rm TeV}$ via the $e^+e^- \to \tilde{\chi}_1^+\tilde{\chi}_1^-$ channel, with a mass reconstruction precision of $\Delta m/m \sim 0.1\%$~\cite{Bai:2021rdg,CLIC:2018fvx}. 

  It is worth emphasizing that the larger mass splittings predicted in our scenario, compared to the highly compressed spectra with typical splittings of several GeV, render the SUSY signals more accessible to detection, as demonstrated by experience in searching for electroweakinos at the LHC~\cite{ATLAS:2021moa}.
   
    \item \textbf{DM Direct Detection} 
    
   Next-generation experiments such as DARWIN aim to achieve unprecedented sensitivities of $3.0 \times 10^{-49}~{\rm cm^2}$ ($2.0 \times 10^{-48}~{\rm cm^2}$) for SI scattering and $4.0 \times 10^{-44}~{\rm cm^2}$ ($3.0 \times 10^{-43}~{\rm cm^2}$)  for SD scattering for DM mass around $40~{\rm GeV}$ ($1~{\rm TeV}$)~\cite{DARWIN:2016hyl}, primarily through increased ton-scale liquid xenon targets and improved spectral discrimination between WIMP signals and backgrounds. Similar sensitivities are expected at the final stage of the LZ experiment~\cite{Mount:2017qzi}. As evident from the right panel of Fig.~\ref{Fig4}, achievement of such sensitivities would enable a complete exploration of the parameter space considered in this study.

    \item {\textbf{DM Indirect detction}}  
    
    Indirect searches for Higgsino DM have been extensively studied, with gamma-ray observatories offering a primary discovery route. The upcoming Cherenkov Telescope Array (CTA) is projected to have sensitivity an order of magnitude beyond current instruments like H.E.S.S~\cite{Rodd:2024qsi}, specifically targeting the monochromatic gamma-ray line from $\tilde{\chi}_1^0 \tilde{\chi}_1^0  \to \gamma\gamma$ annihilation. This provides CTA with potential coverage of the 1 TeV Higgsino parameter space, particularly in scenarios where the Sommerfeld effect, enhanced by a small mass splitting of approximately $\Delta m \sim 1$~MeV, significantly boosts the annihilation rate. In parallel, the next-generation neutrino telescope IceCube-Gen2 will search for high-energy neutrinos from Higgsinos captured and annihilating within the Sun, with expected sensitivity to Higgsino masses up to $|\mu| < 1.5$~TeV, assuming mass splitting of $\Delta m \lesssim 500~{\rm MeV}$~\cite{Rott:2023inm,IceCube-Gen2:2020qha}. Additionally, existing constraints from Fermi-LAT already show a modest 2$\sigma$ excess consistent with a 1~TeV thermal Higgsino, though this result remains subject to large uncertainties in the modeling of the Galactic Center~\cite{dessert2023higgsino}.
     
    It should be noted that these studies have assumed the mass splittings among the Higgsino states less than $500~{\rm MeV}$, substantially smaller than the prediction of our scenario. This difference significantly changes the  Sommerfeld factor and  annihilation cross-sections. A dedicated calculation of differential distributions for these annihilation channels and their comparison with experimental capabilities would be required to fully assess detection prospects in our scenario — such calculations are highly non-trivial~\cite{Beneke:2022eci,Rinchiuso:2020skh} and have not yet been incorporated into automated tools such as the latest version of \textsf{MicrOMEGAs} for the GNMSSM framework~\cite{Alguero:2023zol}. Such an implementation falls beyond the scope of our current investigation.
    
  \begin{figure*}[t]
  \centering
  \resizebox{1.0 \textwidth}{!}{
   \includegraphics{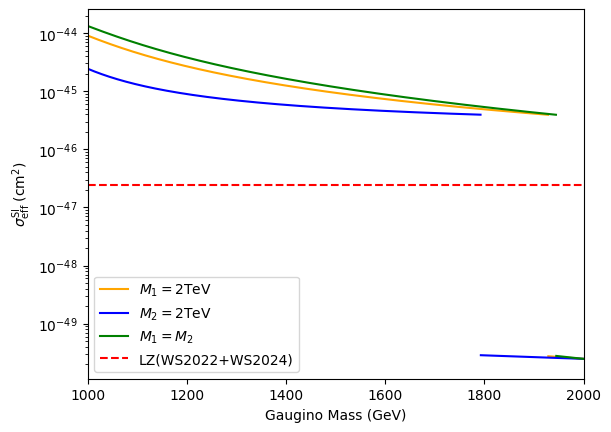}
    \includegraphics{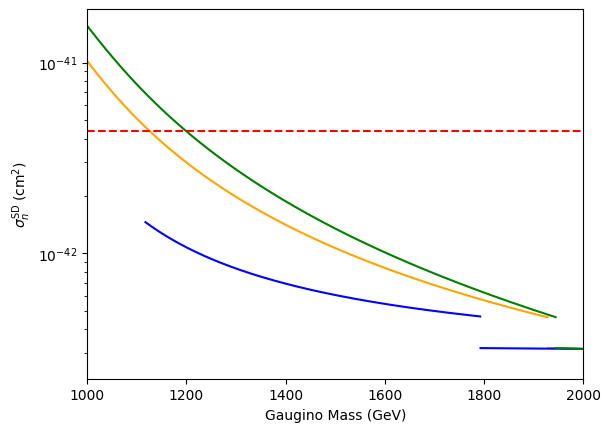}
  }
  \vspace{-0.5cm}
  \caption{Dependence of SI (left panel) and SD (right panel) scattering cross-sections on gaugino mass parameters. The graph illustrates three scenarios: varying $M_2$ while keeping $M_1$ fixed at $2~{\rm TeV}$ (yellow line), varying $M_1$ while maintaining $M_2$ at $2~{\rm TeV}$ (blue line), and simultaneously changing both parameters with $M_1 = M_2$ (green line). All other parameters remain consistent with benchmark point P1. The red lines represent the LZ experimental limits for DM with mass  $|m_{\tilde{\chi}_1^0}| = 865.9~{\rm GeV}$. The discontinuities observed in each curve correspond to points where the DM mass changes sign, a phenomenon detailed in the main text. \label{Fig5}}
\end{figure*}
    
     \item {\textbf{Light Scalar Detection Prospects}}
     
     Our scenario favors a relatively light $h_s$. While direct detection of such a light Higgs at the HL-LHC via ggF/VBF to $\gamma\gamma$ or $WW/ZZ$ faces challenges due to its singlet nature and consequently small production cross-sections, precision measurements of the SM Higgs couplings offer a powerful indirect probe. The substantial mixing between $H_{SM}$ and $Re[S]$ in our scenario makes this approach particularly promising. With projected uncertainties on coupling modifiers reaching 1.8\%~\cite{Mlynarikova:2023bvx}, HL-LHC data, analyzed via tools like HiggsTools \texttt{HiggsTools}~\cite{Bahl:2022igd}, can place strong limits on the singlet sector, potentially exceeding direct search sensitivity~\cite{Heng:2025wng}.
     
\end{itemize}

\begin{figure*}[t]
  \centering
  \resizebox{0.6 \textwidth}{!}{
   \includegraphics{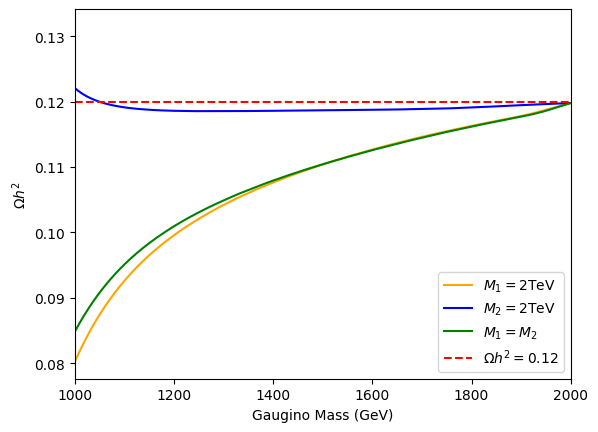}
  }
  \vspace{-0.5cm}
  \caption{Same as Fig.~\ref{Fig5}, but for the dependence of relic abundance on the gaugino mass. \label{Fig6}}
\end{figure*}

\subsection{Gaugino mass dependence}

In the preceding analysis, we have fixed the gaugino masses at $2~{\rm TeV}$ for simplicity. This choice allows us to focus specifically on the novel effects of Higgsino-Singlino mixing. This section explores how variations in gaugino masses affect DM properties, providing a more comprehensive understanding of the parameter space. 

The variation of gaugino masses produces several important physical effects that merit systematic investigation. Lighter $M_1$ and $M_2$ values increase the Bino and Wino fractions within the neutralino mass eigenstates, which typically strengthens the couplings between DM and both the Higgs bosons and the Z boson. These enhanced couplings generally result in larger DM-nucleon scattering cross-sections, affecting the viability of the model under direct detection constraints.  Additionally, as established in MSSM studies~\cite{Hisano:2004pv}, the scattering cross-section expressions exhibit qualitatively different behaviors for positive versus negative DM masses. This feature becomes particularly relevant when varying gaugino masses induces sign changes in the DM mass, leading to discontinuous behavior in the cross-sections at specific transition points. Understanding these transitions is crucial for mapping the viable parameter space.

To quantify the gaugino mass effects systematically, we investigated the behavior of scattering cross-sections as $M_1$ and/or $M_2$ vary within the range of $1000-2000~{\rm GeV}$, while maintaining all other parameters fixed at benchmark point P1. 
Throughout this parameter scan, the magnitude of the DM mass changes slightly, but its sign undergoes a discrete flip from positive to negative. Correspondingly, the dominant component transitions from 
nearly maximal Higgsino–Singlino mixing to an approximately pure Higgsino configuration, as revealed by our numerical results.

Fig.~\ref{Fig5} illustrates the dependence of both SI (left panel) and SD (right panel) scattering cross-sections on gaugino masses. Three distinct scenarios are examined to provide a comprehensive picture: varying $M_2$ while keeping $M_1$ fixed at $2~{\rm TeV}$ (yellow curves), varying $M_1$ while keeping $M_2$ fixed at $2~{\rm TeV}$ (blue curves), and varying both masses simultaneously with the constraint $M_1 = M_2$ (green curves).
The results reveal the following noteworthy features:
\begin{itemize} 
\item As anticipated from theoretical expectations, both scattering cross-sections increase monotonically with decreasing gaugino masses across all scenarios. 
\item The current experimental bounds from LZ are exceeded when $M_2$ falls below $1.93~{\rm TeV}$, $M_1$ drops below $1.79~{\rm TeV}$, or when both masses simultaneously decrease below $1.94~{\rm TeV}$ for the three scenarios, respectively. 
\item These critical threshold values precisely correspond to points where the DM mass changes sign. The discontinuous jumps observed in each curve originate from these sign flips, confirming the theoretical prediction of qualitatively different scattering behaviors for positive versus negative DM masses.
\item Comparing the SI and SD scattering, the former imposes tighter constraints on the gaugino masses. 
\end{itemize}

The thermal relic abundance provides another crucial constraint on the gaugino mass parameter space. Fig.~\ref{Fig6} displays how the relic abundance varies with gaugino masses under the same three scenarios examined above.  The results reveal distinct sensitivities to different gaugino masses.  Decreasing $M_2$ leads to a significant reduction in the thermal relic density $\Omega h^2$, pushing it well below the observed cosmological value of $0.120 \pm 0.001$. This behavior indicates that Wino-like components increasingly dominate the annihilation processes as $M_2$ decreases, enhancing the annihilation cross-section and thereby reducing the relic abundance. In contrast, variations in $M_1$ produce only modest changes in the relic density, suggesting that Bino contributions remain subdominant in the parameter space under consideration. This asymmetric response to $M_1$ and $M_2$ variations provides valuable guidance for parameter space exploration.

With further reduction of gaugino masses beyond the ranges explored here, the DM would transition from being Higgsino-dominated to becoming Bino- or Wino-dominated. Such a fundamental shift in composition would dramatically alter the phenomenology, which is not the focus of this study.  

It should be noted that light Wino-dominated DM faces stringent constraints from indirect detection experiments and LHC experiments. Gamma-ray observations by Fermi-LAT and H.E.S.S. of dwarf spheroidal galaxies and the Galactic Center severely limit such scenarios~\cite{Safdi:2025sfs,Fan:2013faa,Cohen:2013ama,Rinchiuso:2018ajn}. These constraints become particularly severe when the Wino component achieves the correct relic abundance through non-thermal production mechanisms, such as decay of heavier superpartners or moduli. Additionally, LHC searches for electroweakinos have excluded substantial portions of the light Wino parameter space through exploring the signals of disappearing tracks and soft leptons~\cite{ATLAS:2021moa,ATLAS:2021yqv}.

In summary, our systematic analysis demonstrates that the chosen gaugino masses around $2~{\rm TeV}$ represent a phenomenologically viable region that enables exploration of the novel Higgsino-Singlino mixing effects while maintaining consistency with current experimental observations. The systematic variation of gaugino masses reveals clear boundaries beyond which experimental constraints are violated, providing essential guidance for future phenomenological studies in extended supersymmetric models. This analysis confirms that the interesting Higgsino-Singlino mixing physics studied in this work operates within a well-defined and experimentally viable parameter region.

\section{\label{conclusion}Conclusion}

Higgsino DM represents a compelling candidate within supersymmetric frameworks, offering a unique convergence of theoretical insight and experimental motivations. While the MSSM traditionally positioned Higgsino DM at around 1 TeV to achieve the correct relic density, we showed that there still exist viable parameter regions in the GNMSSM with Higgsino DM at significantly lower masses via mixing with Singlino, even confronting with current severe LZ constraints.

We performed a comprehensive analysis of the GNMSSM, focusing on the Higgsino-dominated neutralino as a DM candidate. Our study systematically explored the parameter space by incorporating a range of experimental constraints, including the latest $125~{\rm GeV}$ Higgs data and the LZ results in 2024. Our analysis revealed the following key findings:

\begin{itemize}
  \item \textbf{Lower than 1 TeV Higgsino DM}
   
   In the GNMSSM, Higgsino-dominated neutralinos can achieve the correct DM relic density at significantly lower masses at approximately 670 GeV, which is substantially below the 1 TeV threshold typical in the MSSM. This feature is facilitated by the  Higgsino-Singlino mixing, which can produce correct relic density  without invoking extensive Higgsino-gaugino mixing that would otherwise compromise direct detection compatibility.

  \item \textbf{Lower DM-nucleon scattering rate} 
  
  Despite the increasingly stringent constraints from experiments like LZ, the GNMSSM maintains a considerable viable parameter space. This robustness stems from the Higgsino-Singlino mixing, which can effectively suppress the SI scattering cross-sections.
  
  \item \textbf{Moderate cancellations} 
  
  In order to satisfy the relic density and direct detection constraints, parameter $\lambda$ has to keep at a low value of ${\cal{O}} (0.1)$. This set a lower SI scattering cross-section globally first, then other parameters like $\kappa$ and $V_h^S$ can be tuned to satisfy the direct detection constraints. This reflects cancellations among different terms in the SI amplitude. 

  \item \textbf{Promising future detection prospects}
  
  The mass of Higgsino-like DM in our scenario may be as low as $670~{\rm GeV}$, with mass splittings of more than several GeV from other heavier states. This scenario could be detected in the future through multiple channels: disappearing tracks at the HL-LHC, DM direct detection experiments, or gamma-ray line signatures in CTA observations. Additionally, the singlet scalar could be detected at the HL-LHC through precision Higgs data measurements.

\end{itemize}

In summary, our analysis indicates that the GNMSSM presents a viable alternative to the MSSM, offering refined strategies to tackle persistent challenges in supersymmetric DM models. By incorporating additional degrees of freedom from the singlet sector, we demonstrate how theoretical constraints can be effectively addressed while ensuring phenomenological consistency. Our results underscore the potential of the GNMSSM to provide a compelling framework for Higgsino DM, offering a promising avenue for future experimental searches.

\appendix

\section*{Acknowledgement}
This work is supported by the National Natural Science Foundation of China (NNSFC) under grant No. 12575110 and Natural Science Foundation of Henan Province under Grant No. 232300421217. We thank Dr. Yangle He and Zhiyang Bao for their helpful discussions during this project. The authors also gratefully acknowledge the valuable discussions and insights provided by the members of the China Collaboration of Precision Testing and New Physics(CPTNP).


\bibliographystyle{CitationStyle}
\bibliography{myrefs}
\end{document}